\documentclass[final]{siamart1116}

\usepackage{tiaStyle}
\usepackage{wrapfig}
\usepackage{tikz-cd}
\usetikzlibrary{arrows}
\DeclareMathOperator{\Diag}{Diag}

\newcommand{\bE}[0]{\mathbb{E}}
\newcommand{\bV}[0]{\mathbb{V}\mathrm{ar}}

\newcommand{\R}{\mathbb{R}}

\newcommand{\veckN}{\mathbf{k}_N}
\newcommand{\veckn}{\mathbf{k}_n}
\newcommand{\veccn}{\mathbf{c}_n}

\newcommand{\An}{\mathbf{A}_n}

\newcommand{\C}{\mathbf{C}}
\newcommand{\Cg}{\mathbf{C}_{(g)}}
\newcommand{\CN}{\mathbf{C}_N}
\newcommand{\Cn}{\mathbf{C}_n}

\newcommand{\Deltan}{\boldsymbol{\Delta}_n}
\newcommand{\I}{\mathbf{I}}
\newcommand{\K}{\mathbf{K}}
\newcommand{\KN}{\mathbf{K}_N}
\newcommand{\Kn}{\mathbf{K}_n}
\newcommand{\SN}{\boldsymbol{\Sigma}_N}
\newcommand{\Sn}{\boldsymbol{\Sigma}_n}

\newcommand{\hnug}{\hat{\nu}_{(g)}}
\newcommand{\Un}{\boldsymbol{\Upsilon}_n}

\newcommand{\Lan}{\boldsymbol{\Lambda}_n}

\newcommand{\nuN}{\hat{\nu}_N}

\usepackage[yyyymmdd,hhmmss]{datetime}
\usepackage[colorinlistoftodos,textsize=footnotesize]{todonotes}
\usepackage[normalem]{ulem}

\newcommand{\TheTitle}{Parameter and Uncertainty Estimation for Dynamical Systems Using Surrogate Stochastic Processes}
\newcommand{\ShortTitle}{Inference for Dynamical Systems via Surrogates}
\newcommand{\TheAuthors}{M. Chung, M. Binois, R.B. Gramacy, D.J. Moquin, A.P. Smith, A.M. Smith}
\headers{\ShortTitle}{\TheAuthors}
\title{{\TheTitle}\thanks{Submitted to the editors, \shortdate \today.
\funding{This work was supported by USDA National Institute of Food and Agriculture [grant no. 2016-08687], by NIH National Institute of Allergy and Infectious Diseases [grant nos. AI125324 and AI100946], and by ALSAC. Any opinions, findings, conclusions, or recommendations expressed in this publication are those of the author(s) and do not necessarily reflect the view of these entities. }}}

\author
{%
  Matthias Chung\thanks{Department of Mathematics, Computational Modeling and Data Analytics Devision, Academy of Integrated Science, Virginia Tech, Blacksburg, VA,
  (\email{mcchung@vt.edu}, \url{http://www.math.vt.edu/mcchung}).}
 \and
  Mickaël Binois\thanks{Booth School of Business, The University of Chicago, Chicago, IL, (\email{mickael.binois@chicagobooth.edu}).}
 \and
  Robert B. Gramacy\thanks{Department of Statistics, Virginia Tech, Blacksburg, VA, (\email{rbgramacy@vt.edu}).}
 \and
  David J. Moquin\thanks{Department of Internal Medicine, University of Tennessee Health Science Center, Memphis, TN, (\email{dmoquin@uthsc.edu}).}
 \and
  Amanda P. Smith\thanks{Department of Pediatrics, University of Tennessee Health Science Center, Memphis, TN, (\email{amanda.smith@uthsc.edu}).}
 \and
  Amber M. Smith\thanks{Department of Pediatrics, University of Tennessee Health Science Center, Memphis, TN, (\email{amber.smith@uthsc.edu}).}
}

\usepackage{amsopn}

\begin{document}

\maketitle

\begin{abstract}
Inference on unknown quantities in dynamical systems via observational data is essential for providing meaningful insight, furnishing accurate predictions, enabling robust control, and establishing appropriate designs for future experiments. Merging mathematical theory with empirical measurements in a statistically coherent way is critical and challenges abound, e.g.,: ill-posedness of the parameter estimation problem, proper regularization and incorporation of prior knowledge, and computational limitations on full uncertainty qualification. To address these issues, we propose a new method for learning parameterized dynamical systems from data. In many ways, our proposal turns the canonical framework on its head. We first fit a surrogate stochastic process to observational data, enforcing prior knowledge (e.g., smoothness), and coping with challenging data features like heteroskedasticity, heavy tails and censoring. Then, samples of the stochastic process are used as ``surrogate data'' and point estimates are computed via ordinary point estimation methods in a modular fashion. An attractive feature of this approach is that it is fully Bayesian and simultaneously parallelizable. We demonstrate the advantages of our new approach on a predator prey simulation study and on a real world application involving within-host influenza virus infection data paired with a viral kinetic model.
\end{abstract}



\section{Introduction \& Background}\label{sec:intro}
Standard data fitting for dynamical systems uses a least squares framework in which the Euclidean distance between data and model, a computer-implemented solver, is minimized. Parameter estimation schemes often begin with an initial guess on the values of each coordinate, followed by repeated solving of the dynamical system via numerical integration techniques as directed by a search algorithm, until a termination criteria is satisfied. With many data sets, including the influenza virus infection data discussed here, the optimization problem is ill-posed. As a result, meaningful parameter and uncertainty estimates can remain illusive \cite{engl1996regularization}. Here, we propose a new apparatus, which front-loads with statistical modeling as a means of elevating the potential for sensible uncertainty quantification.

Our proposed method turns the canonical least squares framework on its head. We begin with a continuous, fitted stochastic process to generate feasible data approximations, e.g., smoothness-preserving dynamics. Then, rather than data fitting, we perform sample process fitting. That is, we use a least squares framework to fit samples of the stochastic process (i.e., solution trajectory) rather than fitting the data itself. A major advantage of this framework is that regularization may be introduced more organically in order to promote well-posedness and ultimately lead to more meaningful/useful parameter estimates, as we demonstrate. Another advantage is its intuitive appeal as regards the propagation of uncertainty, via mapping posterior predictive surfaces to posterior distributions on model parameters. Finally, relative to similar alternatives, such as straightforward Bayesian modeling \cite{smith2013uncertainty} or more elaborate Bayesian computer model calibration schemes \cite{kennedy:ohagan:2001,higdon2004combining}, both of which require serial computation via Markov Chain Monte Carlo (MCMC), our methodology is a simple Monte Carlo scheme. Therefore, it offers the potential for vast (and embarrassingly parallel) distributed computation.

Our methodological developments are motivated by experimental data and a parameterized dynamical model describing \textit{in vivo} viral load kinetics during influenza virus infection. As we illustrate, the data present some parameter estimation challenges that can thwart straightforward fitting methods typically used to infer unknown model parameters. One challenge is that the samples from murine infection are destructive and serial sampling of individuals is unavailable to determine viral loads \cite{smith2011effect,smith2018dd}. To address this challenge, data are collected for several animals and at many times after infection. The small heterogeneity in sampling indicates high reproducibility and a robust curve that can be used to extrapolate the average time course of virus dynamics \cite{smith2018dd}. Even so, we often rely on statistically formulated prior beliefs about the dynamics in order to match the data with rigid assumptions posed by the mathematical models. Additional complications arise when the data exhibits input-dependent noise (i.e., they are heteroskedastic), has heavy tails (i.e., they are leptokurtic), and/or there is a lower limit of detection (LOD) within the measurement assay (i.e., censoring). Our aim is to develop an inferential scheme that can cope with such nuances and data deficiencies yet remain computationally tractable and provide full uncertainty quantification via fully Bayesian posterior inference.

Toward that end, the remainder of the paper is organized as follows. We complete our introduction section below by providing background on parameter estimation methods for dynamical systems by way of motivating our new approach, which is briefly outlined before a full treatment in Section~\ref{sec:pe}. Here, we detail the influenza virus infection data, the within-host viral kinetic model, and the challenges associated with heteroskedasticity, outliers, and censored observations to motivate the development a tailored modeling scheme, based on Gaussian processes, in Section~\ref{sec:influenzaGP}. Section~\ref{sec:num} provides detailed empirical work, first via a simulation study as a means of illustration, and then on our motivating influenza example. We conclude with a brief discussion in Section~\ref{sec:discussion}.

\subsection{Solvers and inference}\label{sub:peSolver}
Solving ordinary differential equation (ODE) model-constrained parameter estimation problems may be computationally challenging for various reasons, including having a limited number of observations, high levels of noise in the data, chaotic system dynamics, nonlinear system models, and large numbers of unknown parameters. Many of these challenges appear in biological systems~\cite{Vogel2002,Aster2012}; hence, parameter estimation for dynamical systems for biological applications is of high interest and an active area of research~\cite{Voit2012,Bock1984,Ramsay1996,chung2015robust}.

\begin{figure}[h]
  \begin{boxedminipage}{\textwidth}
  \begin{center}\footnotesize
    \begin{tikzpicture}[>=latex', node distance=30ex]
      \node (p) {$\bfp$};
      \node [right of = p,shape = circle] (y) {$\bfy(\bfp)$};
      \node [right of = y] (c) {$\bfs(\bfy(\bfp))$};
      \node [right of = c] (J) {$\norm[2]{\bfs(\bfy(\bfp))-\bfd}^2 + \calR(\bfp)$};
      \node [above right of = c, node distance=11ex] (d) {$\bfd$};
      \draw [->,ultra thick,blue] (p) -- node[above, black] {parameter-to-model} (y);
      \draw [->,ultra thick,blue] (y) -- node[above, black] {model-to-data} (c);
      \draw [->,ultra thick,blue] (c) -- node[below, black] {loss} (J);
      \draw [->,ultra thick,blue,rounded corners=20pt] (d) |- node[below = 0ex, black] {} (J);
      \draw [->,ultra thick,blue, loop below,rounded corners=30pt, below] (p) -- (0,-1.2)  -- (6,-1.2) node[black, above] {prior knowledge/regularization} --  (11.3,-1.2) --   (11.9,-.2);
    \end{tikzpicture}
  \end{center}
  \caption{Illustration of the ingredients of a classical parameter estimation problem. Parameter $\bfp$ defines the specific model $\bfy$, which then gets mapped by $\bfs$ onto the data $\bfd$, and finally measured with a quality measure $\calJ$, that may include prior knowledge $\calR$ about $\bfp$. For the inverse problem we want to find a $\widehat\bfp$ with best quality measure $\calJ$, given $\bfy$, $\bfs$, regularization $\calR$, and data $\bfd$.}\label{fig:inverseproblem}
  \end{boxedminipage}
\end{figure}
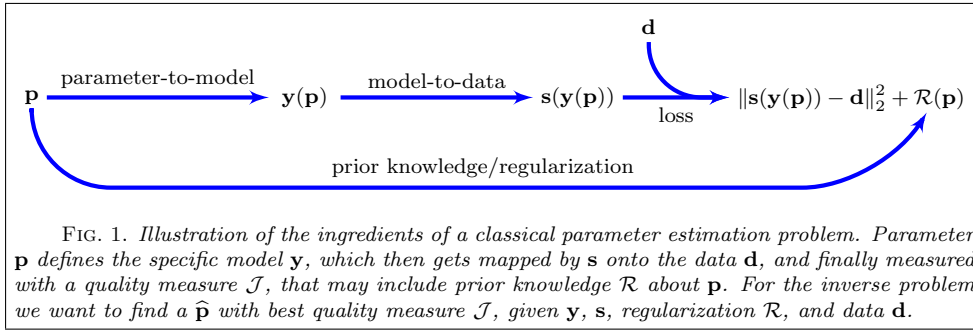

A classic parameter estimation problem may be stated as follows
\begin{gather} \label{eq:pe}
  \min_{\bfp \in \calP} \ \calJ(\bfp) = \norm[2]{\bfs(\bfy(\bfp))-\bfd}^2 + \calR(\bfp) \qquad \mbox{subject to } \bfy' = \bff(t,\bfy,\bfp).
\end{gather}
A diagram of this process is shown in Fig.~\ref{fig:inverseproblem}. Here, $\bfy:\bbR\times \bbR^m \to \bbR^n$ is a solution of a parameter dependent ordinary differential equation (ODE) $\bfy' = \bff(t,\bfy,\bfp)$. We assume that the state solution $\bfy$ is uniquely determined for any $\bfp \in \calP \subset \bbR^m$, where $\calP$ is the feasible parameter space, e.g., $\bfp$ may contain non-negative growth rates $p_j \geq 0$. In biological systems, we often consider initial value problems and may include the initial condition $\bfy(t_0) \in \bbR^n$ in addition to the unknown model parameters, $\bfp$. Throughout this work, we assume that $\bfp$ contains model parameters and initial conditions $\bfy(t_0)$ such that the initial value problem is uniquely determined. The functional $\bfs:\bbR^n \to \bbR^N$ is an operator that maps the state solution $\bfy$ onto an observation space $\calD \subset \bbR^N$, and $\bfd\in\calD$ are the available experimental measurements. Measurements $\bfd$ may only be accessible in a frequency domain or for limited states at discrete times. For example, in predator-prey systems, one may be able to monitor the predator population at certain times while observation of the prey are not available.

In Eq.~\eqref{eq:pe}, we assume that the discrepancy between the model prediction $\bfs(\bfy(\bfp))$ and the data $\bfd$ is given by the Euclidean norm $\norm[2]{\,\cdot\,}$. The methodology we present is not tailored to this choice. Other loss functions, or norms, may be utilized. However, we prefer the $\ell_2$ norm for its computational advantages and familiarity. Prior knowledge on the parameters $\bfp$ may also be included in form of an additive regularization term $\calR(\bfp)$, where regularization prevents ill-posed problems from overfitting the data $\bfd$ (see, for example, \cite{Hansen1998, Aster2012}).

Parameter estimation for the setup described in Fig.~\ref{fig:inverseproblem} and Eq.~\eqref{eq:pe} amounts to solving an {\em inverse problem}. Our inferential scheme is tailored to the setting where one has \emph{repeated} observations $\bfd = [d_1,\ldots, d_n]\t$ of the states given at discrete time points $\bft = [t_1, \ldots, t_n]\t$, as such a setup appears frequently in a diverse set of biological applications (e.g., as in \cite{chung2015robust,Goebel2013,smith2011effect,smith2013kinetics,smith2018dd}). However, even with such regularity in the data, establishing a coherent parameter and uncertainty estimation scheme with underlying dynamical systems for such observations is challenging. Thus, new computationally tractable methodologies could be of great value.

In the literature, various numerical methods have been proposed to solve model constrained optimization problems such as Eq.~\eqref{eq:pe}. Focusing in particular on biological systems, ``single shooting'' approaches are often utilized~\cite{Stoer2002,Bandara2009}. For this strategy, first an initial guess for $\bfp^{(0)}$ is used to numerically solve the initial value problem using numerical ODE solvers like Runge-Kutta and Adams-Bashforth~\cite{Hairer1993,Hairer2010}. Next, the misfit $\calJ(\bfp^{(0)})$ is computed and depending on the optimization strategy (e.g., gradient based strategies or direct search approaches), a new parameter vector $\bfp^{(1)}$ is chosen. Then, in an iterative fashion $\bfp^{(k)}$, $k = 1,\ldots$ is updated until a $\widehat \bfp$ is found subject to pre-determined optimality criteria \cite{Nocedal2006}. If local optimization methods are utilized, globalization is often achieved by Monte Carlo sampling of the search space, i.e., repeated local optimization with random initial guesses~\cite{Conrad2009}. Empirically, the global minimizer $\widehat \bfp$ is chosen from the set of local minimizers obtaining the minimal objective function value. Certainly, other global optimization strategies may be employed, e.g., simulated annealing~\cite{Kirkpatrick1983,smith2018dd}, evolutionary algorithms~\cite{Simon2013}, or particle swarm optimization methods~\cite{Kennedy1995} to name a few.

Besides single shooting methods, which benefit from a straightforward implementation, more sophisticated multiple shooting and principal differential analysis offer the potential of improved robustness of the estimates \cite{Goel2008,Baake1992,Bock1983,Voit2012,Ramsay1996}. We previously investigated these methods \cite{chung2015robust,Chung2012}, but the implementation and computation present unique challenges. Thus, we restrict our attention to single shooting methods here. Nevertheless, each of the methods mentioned above only provides point estimates $\widehat \bfp$, and therefore extra machinery is required in order to quantify uncertainty. One approach in the literature is to deploy local sensitivity analysis \cite{saltelli2008global,marino2008methodology, caracotsios1985sensitivity}, another utilizes bootstraping methods \cite{stoffer1991bootstrapping}. However, Bayesian methods are also gaining traction by using accelerated Markov chain Monte Carlo (MCMC) methods \cite{Calvetti2007, smith2013uncertainty, Tenorio2017}. Despite their advantages of naturally providing uncertainty estimates, MCMC methods in this context can be particularly computationally burdensome because numerical ODE solvers are embedded in accept--reject calculations and the Markov property limits the scope for potential relief via parallelization.

\subsection{A new approach}\label{sub:newapproach}
A statistically sound strategy avoiding ill-posedness\footnote{A problem is ill-posed if a solution does not exist, is not unique, or does not depend continuously on the data, \cite{Hadamard1923}.} of the problem and, therefore, multiple local minima---mitigating the extent to which Monte Carlo search must be utilized---is to introduce prior knowledge in terms of regularization $\calR$. Note, $\calR$ may be derived from the Bayesian framework and correspond to the negative logarithm of the probability density function of the prior distribution of $\bfp$, see \cite{Calvetti2007}. Without proper tuning, however, regularization may bias heavily towards prior knowledge and, thus, parameter estimates are of little value. Moreover, regularization terms, such as standard Tikhonov $\calR(\bfp) = \lambda \norm[2]{\bfp}^2$ with $\lambda \geq 0$, may be inappropriate. This is particularly true for dynamical systems (e.g., biological systems) because parameter values are largely unknown. On the other hand, prior knowledge of state variable dynamics are often readily available. For instance, the dynamics of $\bfy$ may be expected to have certain smoothness (e.g., \textit{in vivo} blood glucose and insulin levels may have a limited decay rates \cite{Goebel2013}). For example, Gong et al.~\cite{gong1998adaptive} proposed regularization terms for dynamical systems, which introduce smoothness on the state variables $\bfy$. However, as mentioned above, tuning associated regularization parameters may be computationally challenging and determining how to propagate uncertainty arising from such procedures is not straightforward.

By way of a potential remedy, we propose to implicitly introduce regularization on the state dynamics $\bfy$ by imposing a surrogate stochastic process on observations from the ``data-generating process''. Then, using sample realizations from the fitted surrogate, we perform inference and estimate uncertainties of $\widehat \bfp$ using standard parameter estimation methods, e.g., via ``single shooting'' with numerical ODE and least squares solvers. This blends prior assumptions on the dynamics with the observed dynamics. Although simple to describe at a high level, the devil is in the details when it comes to diligent application of such a scheme. With the ultimate goal of providing meaningful uncertainty estimates around $\widehat \bfp$, our new method demands three key ingredients that require substantial methodological development: (1) determining scientifically appropriate, yet implementationally pragmatic mechanisms, for including prior knowledge about the biological systems of interest into the surrogate stochastic process; (2) cope with input-dependent noise under potentially heavy-tailed error distributions efficiently, in both statistical and computational senses; and (3) handle censored observations in the data.

Toward that end, we developed extensions to a so-called heteroskedastic Gaussian process ({\tt hetGP}) \cite{Binois2016} to encourage trajectories of a certain shape, handle Student-$t$ noise distributions, and developed an imputation (or data augmentation) scheme to cope with censored observations. With an appropriate surrogate stochastic processes in hand, we illustrate how full posterior inference over unknown parameters to the dynamical system is a straightforward application of ``single shooting'' Monte Carlo, mapping posterior predictive samples into samples of parameters via the ergodic theorem.
 
\section{Problem Setup and Review} \label{sec:pe}
The basic setup of our proposed methodology is as follows. We fit a surrogate to data $\bfd$, comprising of measured observations from a physical (in our
case, biological) process at a discrete/limited number of indices (in our case time). The fitting mechanism incorporates prior information about the
physical processes as a means of regularization. Then, we generate a set of continuous surrogate data $\left\{\bfg_j(t)\right\}_{j= 1}^J$, drawn as
samples from the fitted stochastic process. Each realization of this sample, indexed by $j$, possess the essence of the data, with an added degree of
regularity owing to the prior, and without intrinsic noise, yet possessing all other sources of variability extrinsic to the true data generating mechanism. For each sample $\bfg_j(t)$ we obtain an estimate $\widehat \bfp_j$ by solving the optimization problem
\begin{equation} \label{eq:gpOpt}
 \widehat \bfp_j = \argmin_\bfp \norm[\calL{_2}]{\bfs(\bfy(\bfp)) - \bfg_j}^2\quad \mbox{subject to } \bfy' = \bff(t,\bfy,\bfp), \quad \mbox{for } j = 1,\ldots, J.
\end{equation}
Here, $\norm[\calL_2]{\,\cdot\,}$ denotes the continuous $\calL_2$ norm on a closed interval $[a,b]$.
The set $\left\{\widehat \bfp_j\right\}_{j= 1}^J$ defines a distribution of estimates of the underlying $\bfp_{\rm true}$.

Through the surrogate and subsequent optimization(s), the method filters data from prior and likelihood to posterior distribution on the unknown parameters
$\bfp_{\rm true}$. The model (combining likelihood and prior) comprises of a statistical component via the fitted surrogate, and a mathematical one via
the choice of $\bfs(\bfy(\cdot))$. Note that priors are not being placed directly on $\bfp_{\rm true}$. Prior knowledge about the entire system is
encapsulated in choices made for the underlying stochastic process, and, thereby, transfers into the samples $\bfg_j$ and carries
over to the estimated state variable $\bfy(\cdot,\widehat\bfp_j)$. For instance, if the stochastic process exhibits smooth dynamics in certain areas, $\widehat
\bfp_j$ will be selected for which this feature is prominent in the state variable $\bfy(\cdot,\widehat\bfp_j)$. If samples from the surrogate come from a Bayesian posterior, then the set $\left\{\widehat \bfp_j\right\}_{j= 1}^J$ comprises of samples from the posterior distribution of $\bfp_{\rm true}$ via the ergodic theorem, as the latter optimization step(s) can be interpreted as a deterministic function, $\bfh$, of the surrogate draws, $\widehat \bfp_j = \bfh(\bfg_j)$.

In a way, all of the learning transpires in the first step (i.e., fitting the surrogate data) because this is the only place data observations are involved. Thus, one can argue that the optimization steps (Eq.~\eqref{eq:gpOpt}) amount to post (learning) processing. From an implementation or algorithmic perspective (i.e., first do this, then that), that characterization is prescriptive. This is what provides the Bayesian posterior interpretation described on the posterior for $\widehat{\bfp}$ (see above). However, that description unfairly diminishes the role of those latter steps in the ultimate posterior distribution. Moreover, there is a feedback loop between the prior elements from the surrogate and those from the mathematical modeling components. The former is tasked with producing surrogate data of the form assumed as prerequisites for the latter. Therefore, at least conceptually, the two prior elements are intimately linked. Care is required when choosing appropriate
components for each stage in the process, in particular depending on the nature of the data generating mechanism. Therefore, in what follows, we describe the data $\bfd$, appropriate mathematical models for that process $\bfy$, and choices for appropriate surrogates in our setting---essentially inverting the order of operations described above.

\subsection{Influenza Data} \label{sec:bio}
Influenza viruses are a frequent cause of lower respiratory tract infections and causes over 15 million infections that result in 200,000 hospitalizations each year \cite{thompson2004influenza}. Infections vary in severity from mild to lethal, where the H1N1 strain in the 1918 ``Spanish Flu'' pandemic claimed over 40 million lives. Despite the prevalence of influenza viruses, the interactions between the virus and host remain poorly understood.

\begin{figure}[bt]
 \begin{center}
 \includegraphics[width = \textwidth]{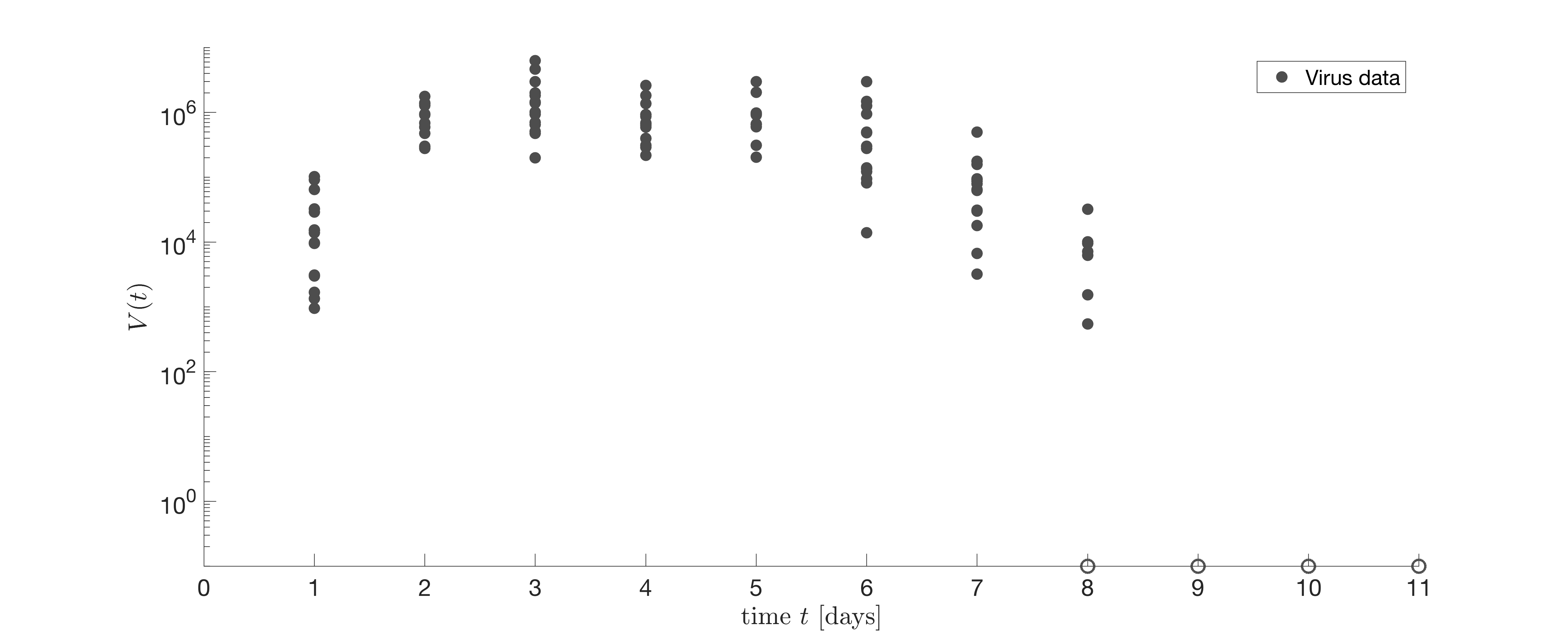}
 \end{center}
		\caption{Viral titers from the lungs of individual mice (black dots) infected with $75\ \mathrm{TCID}_{50}$ influenza A/Puerto Rico/8/34 (H1N1) (PR8) \cite{smith2018dd}. Virus is undetectable in some mice at 8 d post-infection (pi) and for all mice at $t = 9, 10,\ \mathrm{and}\ 11$ d pi, (black circles). These are plotted at $10^{-1}$ $\mathrm{TCID}_{50}$ for visualization.}\label{fig:dataInfluenza.jpg}
\end{figure}

Influenza virus infections are typically acute and self-limiting with short incubation (about 2 days) and infectious periods (typically 4--7 days)~\cite{taubenberger2008piv}. Although the majority of infections are confined to the upper respiratory tract, migration to the lower respiratory tract can result in severe pneumonia. To gain deeper insight into infection mechanisms and the rates of viral growth and decay, mathematical models have been developed and paired with experimental data from humans and animals (e.g., reviewed in \cite{smith2011iav,beauchemin2011rev,smith2016critical}). Parameter estimation remains an important aspect of these studies in order to extract meaningful insight about the infection kinetics. Here, we use one data set and a simple model that accurately describes influenza virus kinetics \cite{smith2018dd} to illustrate the method presented here. The viral load data that we use was obtained from infecting groups of BALB/cJ mice with $75\ \mathrm{TCID}_{50}$ influenza A/Puerto Rico/8/34 (PR8) at time $t = 0$ (time of infection) and measuring viral loads via 50\% tissue culture infectious dose ($\mathrm{TCID}_{50}$) at daily time points $t = 1,\ldots, 11$ \emph{days post-infection} typically abbreviated as ``d pi'',  \cite{smith2018dd}. At each time point (1--11 d pi), samples were collected from 15 individual mice. Thus, the data vector $\bfd \in \bbR^{165}$ comprises the viral load data. The data is shown in Fig.~\ref{fig:dataInfluenza.jpg}. The open circle at 8 d pi may be censored values or true zero values, while the data a 9--11 d pi are thought to be exclusively and reproducibly zeros. For simplicity and consistency, we will treat all zero data points as censored values because the $\mathrm{TCID}_{50}$ assay is typically not sensitive enough to capture low viral loads. We arbitrarily set the LOD at $200\ \mathrm{TCID}_{50}$.

Anticipating the modeling developments (discussed later), observe the following from the figure: The trajectory would appear to be unimodal, which is supported by the biology, although there is no way to ``trace'' this from the raw data. Because the samples are destructive, it is not possible to collect a trajectory of measurements for a single mouse over time. In addition, observe that the data exhibit a degree of heteroskedasticity and/or potentially heavy tailed. This is exemplified by the narrow spread of points nearby time point $t=4\ \mathrm{d}$, versus a wider spread at time $t=1\ \mathrm{d}$ and $t\geq 6\ \mathrm{d}$. The heterogeneity in these time points is true biological heterogeneity and corresponds to the times when virus is rapidly increasing/decreasing \cite{smith2018dd}. Although this is expected, the pattern of spread is at odds with a typical mean--variance relationship (i.e., the spread goes up when mean goes up). Thus, simple transformations are unlikely to help. Finally, it is known that the viral load starts at zero and, therefore, so does the variance \cite{smith2018dd}. This is because virus rapidly infects cells or is cleared in the early stages of infection (0--4 hours post-infection) \cite{smith2018dd}. Similarly, the viral load declines to zero as time increases past 8 d pi. Although these features may easily be accommodated via latent {\em mean} quantities (described later), this exacerbates the heteroskedastic nature of the data and requires more nuanced treatment.

\subsection{Mathematical Modeling} \label{sub:mm}

Because viral dynamics are rapid and complex, studying influenza virus infections with experimental models alone is challenging. Thus, mathematical models have been employed to help identify and detail the mechanisms responsible for controlling viral growth and resolving the infection (reviewed in \cite{smith2011iav,beauchemin2011rev}). These studies have shown that viral load dynamics can be accurately described using 3--4 equations without inclusion of specific innate and adaptive immune responses. Indeed, we recently developed a model that accurately recapitulates the viral load data, including the rapid clearance of virus between 7--9 d pi \cite{smith2018dd}. The model tracks susceptible epithelial (``target'') cells $T$, two classes of infected cells $I_1$ and $I_2$, and virus $V$. In this model, target cells become infected with the virus at rate $\beta V$ per cell. Once infected, these cells enter an eclipse phase $I_1$ at rate $\kappa$ per cell before transitioning to produce the virus at rate $\rho$ per cell $I_2$. Viral loads are cleared at rate $c$ and virus-producing infected cells $I_2$ are cleared in a density dependent manner with maximal rate $\delta/K_d$, where $K_d$ is the half-saturation constant. The following system of differential equations describes these dynamics \cite{smith2018dd}.
\begin{align}\label{eq:im}
 \begin{split}
 T' &= -\beta TV, \\
 I_1' &= \beta TV - \kappa I_1, \\
 I_2' &= \kappa I_1 - \frac{\delta I_2}{K_d + I_2}, \\
 V' &= \rho I_2 - cV.
 \end{split}
\end{align}
The system is of the form $ \bfy'= \bff(t, \bfy, \bfp)$, as considered in Eq.~\eqref{eq:pe}. The state variables are given by $\bfy(t) = [T(t), I_1(t), I_2(t), V(t)]\t$. The parameters $\beta, \kappa, \delta, K_d, \rho, c$ and initial conditions $T(0), I_1(0), I_2(0), V(0)$ uniquely determine the initial value problem. Given an initial condition, this system can be solved numerically with standard ODE solvers, see \cite{Hairer1993,Hairer2010}.

\subsection{Gaussian Process Surrogate Modeling} \label{sec:gp}
Gaussian process (GP) regression, or surrogate modeling, offers a nonparametric framework for estimating functions. GPs are typically trained on a vector of $N$ observations or outputs, $\bfd = [d_1, \dots, d_N]\t$, observed at design of input locations $\bft = [t_1, \dots, t_N]\t \in \bbR^N$. The input and output spaces may be multi-dimensional, however, they both are scalar in the application here. The training data need not be ordered, but as in our discussion here where the inputs are times, we will presume that $t_j \leq t_{j+1}$. A GP is completely defined by its mean and covariance structure, which defines a multivariate normal (MVN) distribution on a finite collection of realizations of the outputs $\bfd$ as a function of inputs $\bft$.

We assume a zero-mean GP prior, which is a common simplifying assumption in the computer simulation modeling literature, e.g., \cite{Santner2013}. This has the effect of moving the modeling effort exclusively into the covariance structure, which is defined by a positive definite kernel $k(\cdot, \cdot)$ function and yields the $N \times N$ covariance matrix of the MVN. The MVN is usually determined by spatial (e.g., Euclidean) distance in the input $t$-space. There are many common choices of kernel functions. The power exponential and Mat\'ern families are the most common. Both of these contain a small number of hyperparameters that are usually learned from the data (for details see \cite{Rasmussen2006,Santner2013}). The specification is completed by choosing a noise process on the output $\bfd$-variables. Typically, this is independent and identically distributed (iid) Gaussian with variance $v(t)$. In some (mainly historical) computer modeling contexts, the simulations are deterministic. In this case, $v(t_j) = 0$. Stochastic simulations are increasingly common, and GP-based surrogate models usually treat the variance function $v$ as constant, which leads to a homoskedastic fit. However, our influenza virus infection example will benefit from a \emph{heteroskedastic} modeling capability in addition to heavier tailed noise distribution---a detail we will return to below. For now, we treat $v(t)$ generically.

The above description can be summarized by the following data-generating specification.
\begin{equation}
\bfd \sim \mathcal{N}(\bfzero, \KN), \quad
\mbox{where} \quad \KN \in \bbR^{N\times N} \mbox{ with} \quad \left(\KN\right)_{ij} = k(t_i, t_j) + \delta_{ij} v(t_i), \label{eq:gp}
\end{equation}
where $\delta_{ij}$ is the Kronecker delta such that the diagonal of $\KN$ is augmented with the ``noise variance''. Any hyperparameters to $k(\cdot, \cdot)$ can be inferred through the likelihood implied by the MVN above, say, via maximum likelihood estimation (MLE). Perhaps the most distinctive feature of this setup is that, due to simple MVN conditioning rules, the predictive distribution at a new input location site $t$ ($d(t) | \bfd$) is (univariate) Gaussian with parameters
\begin{align}
 \mu(t) &= \bE \left( d(t) \mid \bft, \bfd \right) = \veckN(t)^\top \KN^{-1} \bfd \mbox{ and } \nonumber \\
  \sigma^2(t) &= \bV(d(t) \mid \bft, \bfd) = k(t, t) + v(t) - \veckN(t)^\top \KN^{-1} \veckN(t), \mbox{ with } \label{eq:gppred}
 \\
  k_N(t, t') &= \mathbb{C}\mathrm{ov}(d(t), d(t') \mid \bft, \bfd) =
k(t, t') - \veckN(t)^\top \KN^{-1} \veckN(t') + \delta_{t=t'}v(t), \label{eq:wpred} \nonumber
\end{align}
where $\veckN(t) = [k(t, t_1), \dots, k(t, t_N)]^\top$. Vectorized versions, essentially tabulating the covariance structure described above into a full MVN structure, generalize these equations to a joint predictive distribution over a set $\mathcal{T}$ of new locations. One disadvantage, which is apparent from inspecting the equations above, is that the method can be computationally (and storage) intensive in the presence of moderately large data sizes (large $N$), owing to the cubic cost of matrix decomposition and the quadratic cost of storage for $\KN$. Similar bottlenecks are in play when working with the likelihood for hyperparameter inference.

Excluding the deterministic case (i.e., using $v(t_j) = 0$), having replicate $d_i$ and $d_j$ observations at the same input locations, $t_i = t_j$, can be useful for separating signal from noise \cite{Binois2016}. Our influenza virus infection data naturally has this feature. It turns out that having replication in the design also yields computational advantages \cite{Binois2016}. Let $\bar t_i$, $1= i, \dots, n$ the $n \leq N$ \emph{unique} input locations, and $d_i^{(j)}$ be the $j^\mathrm{th}$ out of $a_i \ge 1$ replicates, i.e., $j=1,\dots, a_i$, observed at $\bar t_i$, where $\sum_{i = 1}^n a_i = N$. Also, let $\bar \bfd = [\bar d_1, \dots, \bar d_n]\t$ stand for averages over replicates, $\bar d_i = \frac{1}{a_i}\sum_{j =1}^{a_i} d_i^{(j)}$. Then, it can be shown \cite{Binois2016} that predictive equations (Eq.~\eqref{eq:gppred}) may be applied with $\bar\bfd$ in place of $\bfd$ and unique-$n$ matrices and vectors in place of full-$N$ ones using $\veckn(t) = [k(t, \bar t_1), \dots, k(t, \bar t_n)]\t$ and $(\Kn)_{ij} = k(\bar t_i, \bar t_j) + \delta_{ij} v(\bar t_i) / a_i$. In effect, this unique-$n$ scheme is utilizing an $\mathcal{O}(n)$ number of sufficient statistics for $\mathcal{O}(N)$ degrees of freedom.

\section{Influenza Surrogate} \label{sec:influenzaGP}
The standard GP setup falls short in the context of our motivating influenza virus infection data provided in Fig.~\ref{fig:dataInfluenza.jpg}. Although replications are well handled by sufficient statistics, the variance function is unknown and may be changing throughout the input space. Furthermore, the tails may be heavier than Gaussian, and a portion of the data fall below the LOD and are censored. In the following, we provide extensions to remedy these shortcomings: first, we leverage recent advances in heteroskedastic regression (Section~\ref{sub:hetGP}); second, we present a novel approach to Student-$t$ errors in the heteroskedastic GP context (Section~\ref{sub:heavyhetGP}). Finally, we develop a data-augmentation scheme for handling censored observations in Section~\ref{sec:censor}.

\subsection{Heteroskedastic GP Modeling}\label{sub:hetGP}
In practice, $v(\cdot)$ is seldomly known and the noise variance must be estimated from data, as with any other unknown quantity with the exception of \cite{Picheny:2013}. A simple, yet effective, way of estimating the variance from data in a GP setting under replication is to use empirical, or moment-based, estimators. That is, select the diagonal matrix $\{\delta_{ij} v(t_i)\}$ in Eq.~\eqref{eq:gp}) as
\begin{equation}
 \widehat{\boldsymbol{\Sigma}}_n = \diag{\hat{\sigma}^2_1/a_1, \dots, \hat{\sigma}^2_n/a_n}, \quad \mbox{ where } \quad \hat{\sigma}^2_i = \frac{1}{a_i - 1} \sum \limits_{j = 1}^{a_i} (d_i^{(j)} - \bar d_i)^2.
 \label{eq:sigmahat}
\end{equation}
The resulting predictor is known in the literature as stochastic kriging \cite{Ankenman2010}, or SK. SK has the benefit of accommodating heteroskedastic data, as each input's variance is estimated independently of the rest. It is also computationally advantageous due to working with $n$ rather than $N$ quantities. However, two disadvantages are: (i) the method requires a minimum amount of replication ($a_i \gg 10$ is recommended) both for stability and for its asymptotic properties; and, perhaps more importantly for our needs, (ii) it yields no direct estimate of $v(t)$ out-of-sample, i.e., for a $t$ not in the training design $\bft$. For the latter, it is recommended to fit a second independent GP to the logarithm of the empirical variances $(\hat{\sigma}^2_1/a_1, \dots, \hat{\sigma}^2_n/a_n)$.

If two processes, one for the mean and another for the variance, are to be fit from the same data, then ideally the inference for those two unknowns would be performed jointly. Such approaches pre-date SK in the machine learning literature \cite{goldberg:williams:bishop:1998}, where the ($\log$) variances are treated as latent variables under a GP prior. Here, inference requires cumbersome MCMC calculations over all $N$ unknowns, and implies a complexity of $\calO(N^4)$, which is prohibitive even for modest $N$. Subsequently, researchers replaced the MCMC with point-based alternatives, e.g., using expectation maximization (EM), and related methods \cite{kersting:etal:2007,Boukouvalas2014}. However, none of these methods leveraged the computational savings that comes from having replicates in the design as in the case of SK.

A new method called {\tt hetGP}, detailed in \cite{Binois2016}, offers a hybrid between SK and a joint modeling approach. For a stationary kernel, such as the ones mentioned above, we may equivalently write $k(t, t') = \nu c(t - t')$ such that $\K = \nu(\C + \Lan)$ with $(\bfC)_{ij} = c(\bar t_i - \bar t_j)$ and $\log \Lan = \Cg (\Cg + g \An^{-1})^{-1} \Deltan$, where $\Cg$ is the analog of $\C$ for the second GP with kernel $k_{(g)}$ and $\An = \mathrm{diag}(a_1, \dots, a_n)$. That is, $\log \Lan$ is the prediction given by a GP based on latent variables $\Deltan = (\delta_1, \dots, \delta_n)$ that can be learned as additional hyperparameters alongside the second GP hyperparameters of $k_{(g)}$ and its noise $g$. Using this notation, the predictive equations Eq.~\eqref{eq:wpred}) can be represented as
\begin{align}
 \begin{split}
 \mu_n(t) & = \veccn(t)^\top \left(\Cn + \Lan \An^{-1}\right)^{-1} \bar\bfd, \label{eq:wood-gp} \\
 \sigma_n^2(t) &= \nu \left(1 - \veccn(t)^\top \left(\Cn + \Lan \An^{-1}\right)^{-1} \veccn(t) \right).
 \end{split}
\end{align}
Unknown quantities may be inferred via maximum likelihood as follows. The MLE of $\nu$ is
\begin{equation}
\nuN = N^{-1} \left(N^{-1} \sum_{i=1}^n
\frac{a_i}{\lambda_i} s_i^2 + \bar \bfd\t (\C + \An^{-1}\Lan)^{-1} \bar \bfd
\right),
\end{equation}
with $s_i^2 = \frac{1}{a_i} \sum_{j=1}^{a_i} (d_i^{(j)} - \bar d_i)^2$. The remaining hyperparameters can be optimized using the concentrated joint log-likelihood
\begin{align*}
 \log \tilde L =~& - \frac{N}{2} \log \nuN \nonumber - \frac{1}{2} \sum\limits_{i=1}^n \left[(a_i - 1)\log \lambda_i + \log a_i \right] - \frac{1}{2} \log |\C + \An^{-1}\Deltan| \\
 & - \frac{n}{2} \log \hnug - \frac{1}{2} \log |\Cg + g \An^{-1}| + \mbox{const}, \label{eq:jllik}
\end{align*}
with $\hnug = n^{-1} \Deltan^\top (\C + g \An^{-1})^{-1} \Deltan$. Closed form expressions of the derivatives are given in \cite{Binois2016}. Besides being able to cope with input-dependent noise, the coupling of the processes means that no minimum amount of replication is required to perform inference and rely on the resulting predictions. This is in contrast to SK.

To provide an illustration of this method in a more controlled setting---we shall return to the motivating influenza example shortly---consider the motorcycle accident data \cite{silv:1985}, a stochastic simulation modeling the acceleration of the helmet of a motorcycle rider as a function of time just before and after an impact. It possesses both of the features targeted by the method above: light replication and input-dependent noise.
\begin{figure}[ht!]
\centering
\includegraphics[width= \textwidth]{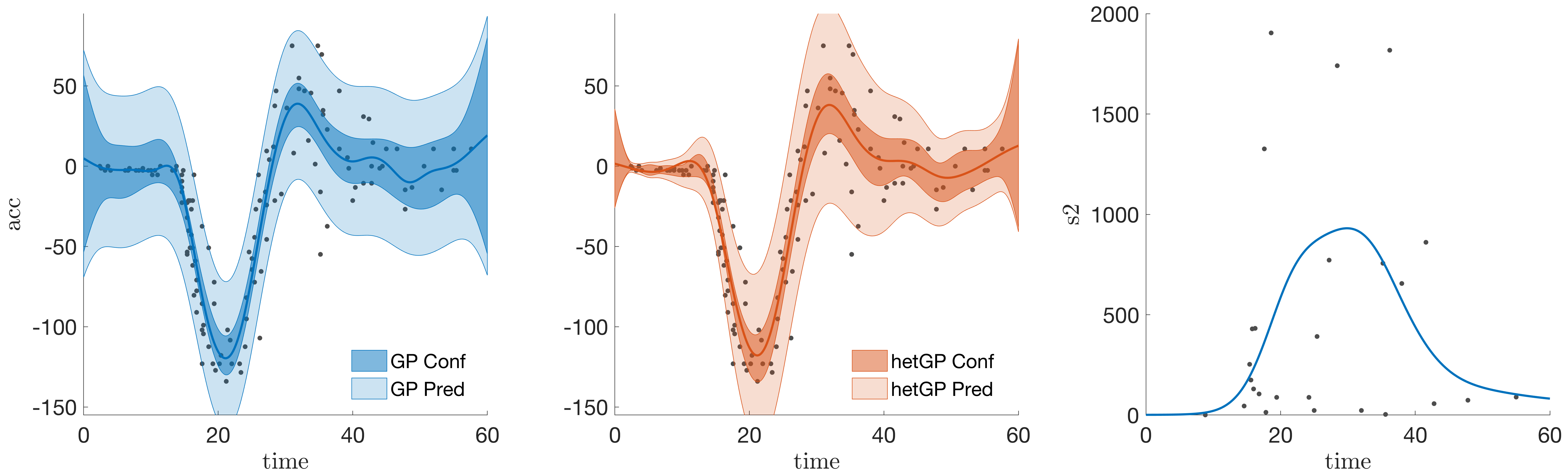}
\vspace{-0.5cm}
\caption{The left panel shows an ordinary GP prediction, whereas the middle panel shows the {\tt hetGP} alternative. Mean (dark) and total (light $+$ dark) are shown. The right panel illustrates the estimated variances (s2 $\equiv \sigma_N^2(x)$) from the heteroskedastic fit (black line), versus time, against the empirical variance estimates for those inputs with two or more replicates.}
\label{f:moto}
\end{figure}
The left panel of Fig.~\ref{f:moto} shows the predictive surface from an ordinary GP fit (Eq.~\eqref{eq:gppred}) while the middle panel shows equivalently the predictive surface for {\tt hetGP}, accommodating heteroskedasticity Eq.~\eqref{eq:wood-gp}). We observe how the former is unable to capture the noise dynamics whereas the latter captures the data better. The mean fits are similar but not identical. The right panel in Fig.~\ref{f:moto} shows the estimate of the noise level more directly, via estimates of the latent $\Deltan$ values. The dots in this panel show the empirical variances for the input locations which have two or more replicates. Basing an estimate of spatial variance on these values (i.e., following SK) would clearly leave something to be desired. For starters, these values are ``choppy'' over time---a feature not evident in a cursory inspection the data. The {\tt hetGP} method furnishes a smooth alternative within a unified mean--variance stochastic modeling framework without the SK requirement of a high degree of replication: all $a_i \gg 10$.

\subsection{Heavy-tailed Heteroskedastic GP} \label{sub:heavyhetGP}
Sometimes the Gaussianity assumption can be overly rigid, particularly in the presence of heavy-tailed perturbations or outliers. Recall that the influenza virus infection data in Fig.~\ref{fig:dataInfluenza.jpg} exhibit this feature. In this setting, Shah et al.~\cite{shah:wilson:ghahramani:2014} showed that Student-$t$ processes (TPs) generalize GPs and share most of their practical appeal. The trick to overcome limitations from previous attempts \cite{Rasmussen2006} is to augment an existing covariance kernel $k(\cdot, \cdot)$ with white noise. Likelihood and predictive equations remain tractable with simple adjustments to the usual closed form expressions. After specifying the degrees-of-freedom parameter $\alpha \in \R_+ \setminus [0,2]$ 
\begin{align} \label{eq:tppred}
 \begin{split}
\alpha_N &= \alpha + N, \\
\mu(t) &= \bE \left( d(t) \mid \bft, \bfd \right) = \veckN(t)^\top \KN^{-1} \bfd,\\
 \sigma^2(t) &= \bV(d(t) \mid \bft, \bfd) = \frac{\alpha + \beta - 2}{\alpha + N - 2} \left(k(t, t) - \veckN(t)\t \KN^{-1} \veckN(t)\right) + v(t), \mbox{ and}\\
 \mathbb{C}\mathrm{o}&\mathrm{v}(d(t), d(t') \mid \bft, \bfd) = \frac{\alpha + \beta - 2}{\alpha + N - 2} \left(k(t, t') - \veckN(t)^\top \KN^{-1} \veckN(t') \right) +\delta_{t = t'} v(t),
 \end{split}
\end{align}
with $\beta = \bfd\t \KN^{-1}\bfd$. Note that the predictive covariance depends on the observed $\bfd$ values. This is in contrast to the Gaussian case Eq.~\eqref{eq:gppred}.

The corresponding log-likelihood is then given by
\begin{align}
 \begin{split}
\log(L) =& -\frac{N}{2} \log((\alpha - 2) \pi) - \frac{1}{2}\log(|\KN|) \\
& + \log \left(\frac{\Gamma \left(\frac{\alpha+N}{2}\right)}{\Gamma \left(\frac{\alpha}{2} \right)} \right) - \frac{(\alpha + N)}{2} \log \left( 1 + \frac{\beta}{\alpha - 2} \right), \nonumber
 \end{split}
\end{align}
where $\Gamma$ denotes Gamma function. In the presence of replicates, it is possible to show that the full-$N$ equations can be expressed by unique-$n$ analogs via the following expressions,
\begin{align*}
\beta = \bfd\t(\tau^2 \CN + \SN)^{-1}\bfd &= \bfd\t \SN^{-1} \bfd - \bar \bfd\t \An \Sn^{-1}\bar \bfd + \bar \bfd\t (\tau^2 \Cn + \An^{-1} \Sn)^{-1} \bar \bfd \\
\log |\KN| = \log |\tau^2\CN + \SN| &= \log |\tau^2 \Cn + \An^{-1}\Sn| + \log |\SN| - \log |\An^{-1} \Sn|.
\end{align*}
An SK-style moment-based estimator of the variance calculated from replicates can be used in these equations to cope with heteroskedasticity. However, this suffers the same drawbacks as in the GP case. For instance, if there are not sufficient replicates, then the result is highly unstable. Additionally, in the TP setting $\alpha$ affects both mean and noise covariances, which further complicates inference and prediction schemes. Fortunately, input dependent (log) noise can be learned via a latent GP in exactly the same way as in {\tt hetGP}, leading to an effective {\tt hetTP} formulation. In fact, it is remarkable that, at least from an implementation perspective, no further modifications are required.

Although one could potentially entertain a TP on the noises, we have not found any practical value for such a setup within our own experimentation. Based on expressions given in \cite{shah:wilson:ghahramani:2014}, closed form derivatives for the log-likelihood are available, similar to \cite{Binois2016}. As these represent a substantial component of our contribution, we provide such expressions in detail in our Appendix~\ref{ap:TP}. An implementation for this {\tt hetTP} is provided as an alternative in our {\tt hetGP} package on CRAN \cite{Binois2017}. To the best of our knowledge, ours is the first application of input dependent, and simultaneously leptokurtic noise in GP regression.

\begin{figure}[ht!]
\centering
\includegraphics[width = \textwidth]{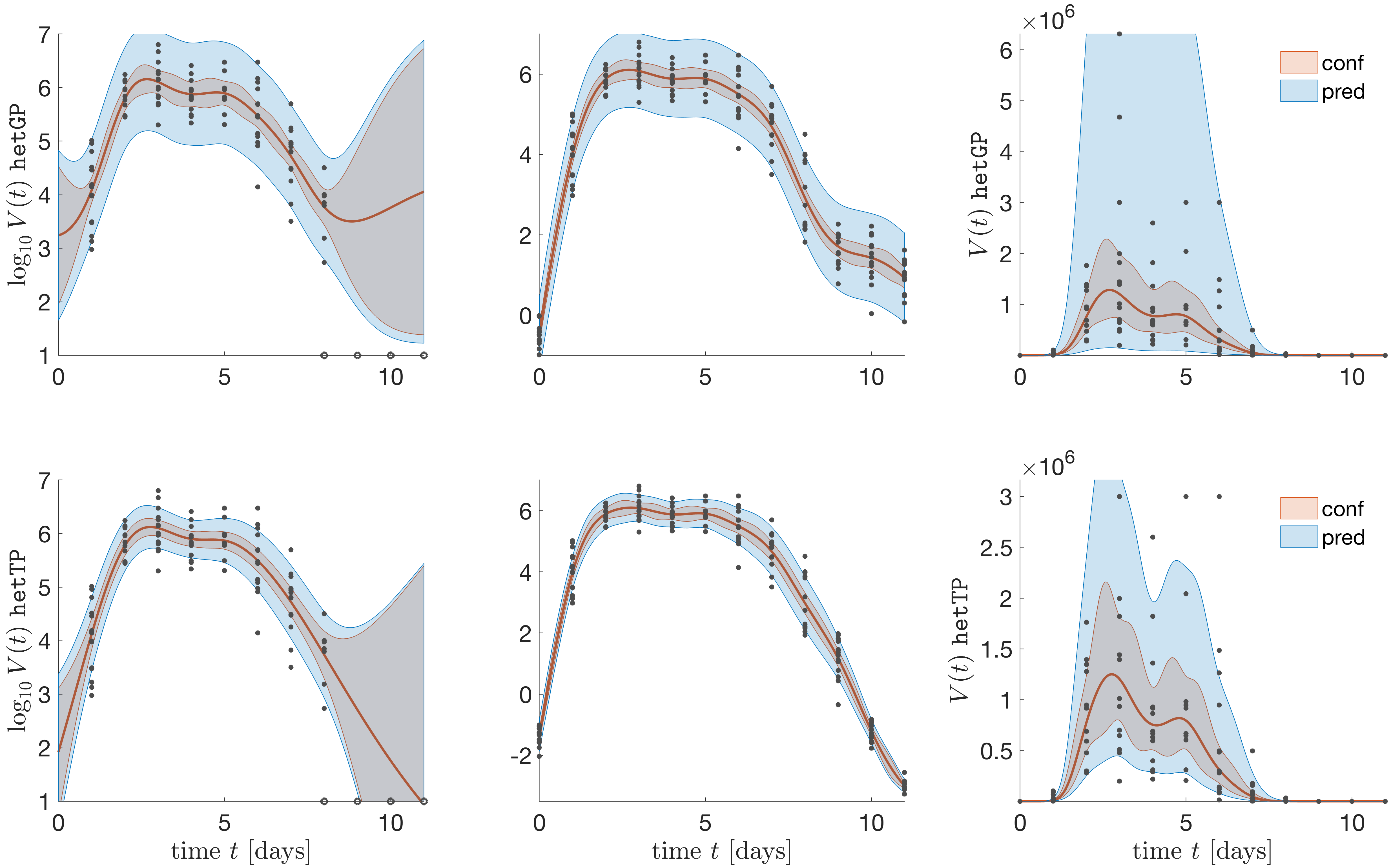}
\vspace{-0.25cm}
\caption{Initial and final models. Top: {\tt hetGP} models. Bottom: {\tt hetTP} models. Left: before data augmentation, $\log_{10}$ scale. Center after data augmentation, $\log_{10}$ scale. Right: after data augmentation, original scale. 95\% confidence and prediction intervals are given as shaded areas, respectively, and the mean is represented as a red curve.}
\label{f:tillus}
\end{figure}

Fig.~\ref{f:tillus} provides an illustration on our motivating influenza virus infection data. The left panel shows a fit to the completely observed portion of that data on the $\log_{10}$ scale, using {\tt hetGP} (top) and {\tt hetTP} (bottom). Observe the narrower intervals provided by the latter, which attributes a larger portion of the noisy observations to outlying events. The middle panel incorporates our data augmentation scheme (described below) for handling censored observations at the extremes of time. Observe that the {\tt hetTP} variation (bottom) provides a ``tighter'' distribution on those censored values and offers a more consistent decline as time passes from 9--11 d pi. The right panel in Fig.~\ref{f:tillus} shows the same pair of plots on a linear scale. In this view, the advantage of treating the largest values, from times 3--6 d pi, as outliers is more readily apparent than it is on the $\log_{10}$ scale. It is the appropriate handling of these output extremes, as noise rather than as signal on the $y$-axis, that can lead to the favorable properties of the output at the beginning and end (i.e., rapid increase at the start of the viral titer dynamics, and the rapid decrease at the end).

\subsection{Censored Observations at the Extremes} \label{sec:censor}
Another feature present in our motivating influenza example are the censored observations at latter times, and a prior for decreasing functions as the data fall into this censoring regime, $t\rightarrow 0$ and $t\rightarrow \infty$. It is known that the virus count ultimately decays to exactly zero \cite{smith2018dd}, however, this presents challenges in $\log_{10}$ space as it corresponds to negative infinity. Therefore, even when the crude option of replacing the censored values with a choice of $\varepsilon \ll 1$ is possible, it creates stochastic modeling challenges, i.e., via GPs (ordinary, {\tt hetGP} and {\tt hetTP}). Subsequent optimization of ODE solutions, based on surrogate data coming out of such fits, exhibit bias in the parameter estimates because the resulting trajectories (e.g., in viral load over time) struggle to reproduce the plateaus that arise.

Therefore, we advocate a softer approach: encouraging a decreasing mean in the GP predictions as $t$ decreases to 0 d pi, starting at 1 d pi, and as $t$ increases to $\infty$, starting at 8 d pi. {\em a posteriori}, such a prior is easy to implement via a rejecting sampling-based data augmentation scheme when obtaining draws from the Gaussian (or Student-$t$) predictive distribution. In what follows, we describe our scheme for data augmentation for censored values as $t$ gets large. The scheme for dealing with $t\rightarrow 0$ proceeds similarly.

The first step involves estimating hyperparameters of the GP covariance kernel, $k(\cdot, \cdot)$, which we obtain via the complete data log-likelihood. Conditional on those settings and on the complete data, we may draw from the predictive equations (Eq.~\eqref{eq:tppred}). Two examples are shown in the left column of Fig.~\ref{f:tillus}. Using those equations, we developed an imputation scheme that proceeds iteratively from the smallest time index with a censored observation, say index $j$ at time $t_j$. In our influenza example in Fig.~\ref{fig:dataInfluenza.jpg}, this is time $t_j=8\ \mathrm{d}$. We then repeatedly draw from the {\em noise-free} predictive equations (e.g., using $v(\cdot) = 0$ in Eq.~\eqref{eq:tppred}, using a set of inputs $\mathcal{T} = \{t_i : t_i \leq t_j\}$ and stop when a draw is obtained that is monotonically decreasing in all censored time indices. At this point, only checking at $t_j$ and $t_{j-1}$ is required. Then, we take a number of draws equaling the number of censored observations from the appropriate noise distribution, conditional on those values being below the censoring threshold. For our motivating influenza problem, this value is $\log_{10}\ 201\ \mathrm{TCID}_{50}$. In the case of {\tt hetGP}, the value is chosen from the Gaussian distribution with variance $\hat{v}(t_j)$. For {\tt hetTP}, a Student-$t$ with degrees of freedom $\alpha_N$ is used and multiplied by $\sqrt{\hat{v}(t_j)}$. Finally, we treat the sampled draws at $t_j$ as actual data values and repeat, moving on to $t_{j+1}$.

\begin{figure}[ht!]
\centering
\includegraphics[width = \textwidth]{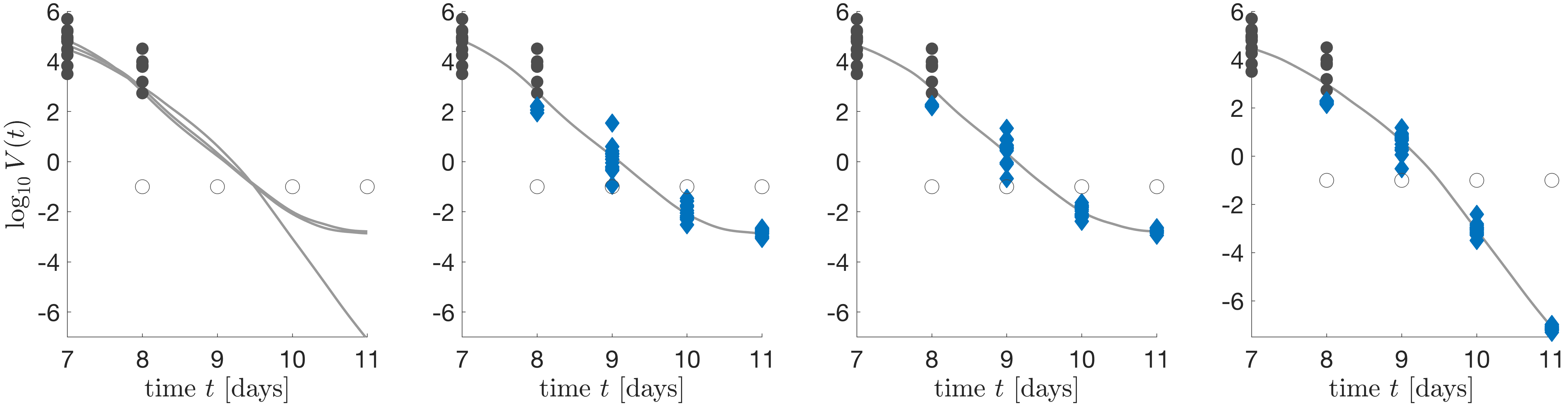}
\vspace{-0.25cm}
\caption{The left-most panel shows three sample paths in gray, overlaid onto a zoomed-in portion of Fig.~\ref{fig:dataInfluenza.jpg} covering time $t \geq 7$. The subsequent three panels show each of those sample paths separately, accompanied by the latent samples (as blue diamond characters) generated for the censored values in the data.}
\label{f:latent}
\end{figure}

After the censored observations have been replaced with ``synthetic'' samples in this way, we obtain draws from the full predictive distribution for the use in the wider parameter estimation exercise (Section~\ref{sec:pe}). Specific illustrations are included below. In this way the scheme is a variation on so-called data augmentation for Bayesian spatial modeling \cite{fridley:dixon:2007}. Fig.~\ref{f:tillus} provides an illustration of the overall surfaces which arise under this scheme, using Gaussian and Student-$t$ innovations respectfully. However, to reduce clutter no actual sample paths or latent samples are shown in that figure. A glimpse at those for {\tt hetTP} is provided in Fig.~\ref{f:latent}, whose panels ``zoom in'' on times 7--11 d pi. The left-most panel shows three draws from the posterior predictive distribution, each of which is conditioned on a separate sample of latent values obtained via the scheme described above. The subsequent three panels in the figure re-draw those sample paths separately, along with a visualization of the latent values which aided in its production.

These surrogate draws (gray lines in Fig.~\ref{f:latent}) have a distribution characterized by the bottom-right panel in Fig.~\ref{f:tillus}, in the case of the {\tt hetTP} model transformed back onto the original scale of the data. Each draw is a sample from a posterior (predictive) distribution. When treated as surrogate data in a ``single shooting'' least-squares search for parameters $\widehat{\bfp}_j$, we obtain a map from surrogate posterior to posterior over parameters to the ODE Eq.~\eqref{eq:im}). These details are laid out algorithmically and illustrated empirically on a toy example and by our motivating influenza example in the following section.

\section{Numerical Experiment} \label{sec:num}
Here, we provide numerical illustrations of the scheme obtained by chaining together the methodological pieces detailed above. Algorithm~\ref{alg:pe} provides a skeleton for the overall procedure. The following discusses the Algorithm detail specification and subroutines with reference to particular procedures and equations provided earlier.
\algrenewcommand\algorithmicrequire{\textbf{input:}}
\algrenewcommand\algorithmicensure{\textbf{output:}}
\algrenewcommand\algorithmicfor{\textbf{parallel for}}
\begin{algorithm} \caption{Parameter \& Uncertainty Estimation via Stochastic Processes}\label{alg:pe}
 \begin{algorithmic}[1]
  \Require data $\bfd$ and ODE model
  \State use $\bfd$ to fit the stochastic process $\calG$
  \For{$j = 1$ to $J$}
   \State generate sample $\bfg_j$ from $\calG$
   \State compute $\widehat\bfp_j$ from Eq.~\eqref{eq:gpOpt} using $\bfg_j$
  \EndFor
  \Ensure $\left\{\widehat\bfp_j\right\}_{j = 1}^{J}$
 \end{algorithmic}
\end{algorithm}
\begin{enumerate}
 \item Inputs include observations $\bfd$ (e.g., data represented in Fig.~\ref{fig:dataInfluenza.jpg} in Section~\ref{sec:bio}) and model equations (e.g., Eq.~\eqref{eq:im} in Section~\ref{sub:mm}).
 \item An appropriate stochastic process $\calG$ needs to be determined, reflecting the data $\bfd$, and given prior knowledge on the dynamics of the system (line~1). Examples are provided in Section~\ref{sec:gp} and Sections~\ref{sub:hetGP}--~\ref{sec:censor}, including variations on GPs (i.e., ordinary, {\tt hetGP}, and {\tt hetTP}).
 \item Monte Carlo samples $\bfg_j$ are drawn from the posterior predictive distribution provided by $\calG$ in line~3. This is typically a straight forward process, however, standard sampling methods (e.g., rejection sampling) may be required to obtain a sample $\bfg_j$ from $\calG$. We follow the discretize-then-optimize approach, discretizing the $\calL_2$ norm in Eq.~\eqref{eq:gpOpt} using a equidistant grid. One option is to match the inputs with the data inputs, i.e., the times at which observations were collected. However, finer or coarser resolutions may be used. We choose a finer resolution, spanning the original range of times imposing the smoothness of the state variables.
 \item Given the sample $\bfg_j$, an optimization scheme and a numerical ODE solver are required to compute point estimates $\widehat\bfp_j$ as discussed in Section~\ref{sub:peSolver} and at the beginning of Section~\ref{sec:pe}. The optimization scheme requires an initial guess $\bfp_0$ and may be chosen differently for each $j$ if desired.
 \item The algorithm returns a set $\left\{\widehat\bfp_j\right\}_{j = 1}^{J}$ reflecting the posterior distribution of parameter estimates.
 \item Although the pseudo-code shows surrogate data generation and parameter estimation happening within the same (potentially parallel) {\bf for} loop, those two steps need not be executed in tandem. After fitting $\calG$, generating a collection of realizations $\bfg_j$, subsequent  fitting of $\widehat{\bfp}_j \mid \bfg_j$, may even be performed offline or on an ad hoc basis.
  Fitting of $\calG$ is the only point of contact between them as regards statistical inference, and therefore this step is independent of the model equations. Other models can be entertained {\em ex post} without revisiting the data or fitting of $\calG$.
\end{enumerate}

\medskip

As a benchmark, we consider a Bayesian approach to parameter inference \cite{smith2013uncertainty} in this setting as applied to our motivating influenza example. The essence of that scheme is a Gaussian likelihood measuring the distance between the data and single shooting paths derived from the ODE under parameters, $\bfp$. Specifically, $\pi(\bfd|\bfp) \propto \exp\left(\hf\norm[2]{\bfs(\bfy(\bfp))-\bfd}^2\right)$. This is paired with independent priors on the individual parameters, often chosen to be uniform in an appropriate range. Our particular choices are application dependent and are detailed below. Inference is facilitated by a Metropolis MCMC \cite{hastings1970monte}. This approach is beneficial in terms of simplicity and is straightforward to implement. However, tuning the Metropolis proposals to obtain adequate mixing of the Markov chains can be highly application dependent, and it is not easily parallelizable. Although such challenges are surmountable, a deficiency that remains is that it is not straightforward to incorporate prior information on the regularity of the observation trajectory, such as smoothness or (local) monotonicity. As we show, this results in a far more diffuse posterior distribution compared to our proposed method.

We turn now to two numerical investigations of our proposed method. For validation, we first investigate our method on a simulation study, and then turn to our motivating influenza problem (Sections~\ref{sec:bio}--\ref{sub:mm}).

\subsection{Simulation Study} \label{sub:simstudy}
In the first simulation study, we assume observations of predator and prey are given. These data are generated using the Lotka-Volterra system
\begin{align}\label{eq:lv}
 \begin{split}
  y_1' &= -y_1 + \alpha_1 y_1 y_2,\\
  y_2' &= y_2 - \alpha_2 y_1 y_2,
 \end{split}
\end{align}
with $\bfalpha_{\rm true} = [1,1]\t$ and initial condition $\bfy_{\rm true}(0) = [y_1(0),y_2(0)]\t = [2, 1/2]\t$ on the interval $t \in [0,10]$ at 20 equidistant time points $t_j = \tfrac{10}{19}(j-1)$, $j = 1,\ldots, 20$.

\begin{figure}[ht!]
 \begin{center}
   \includegraphics[width=\textwidth]{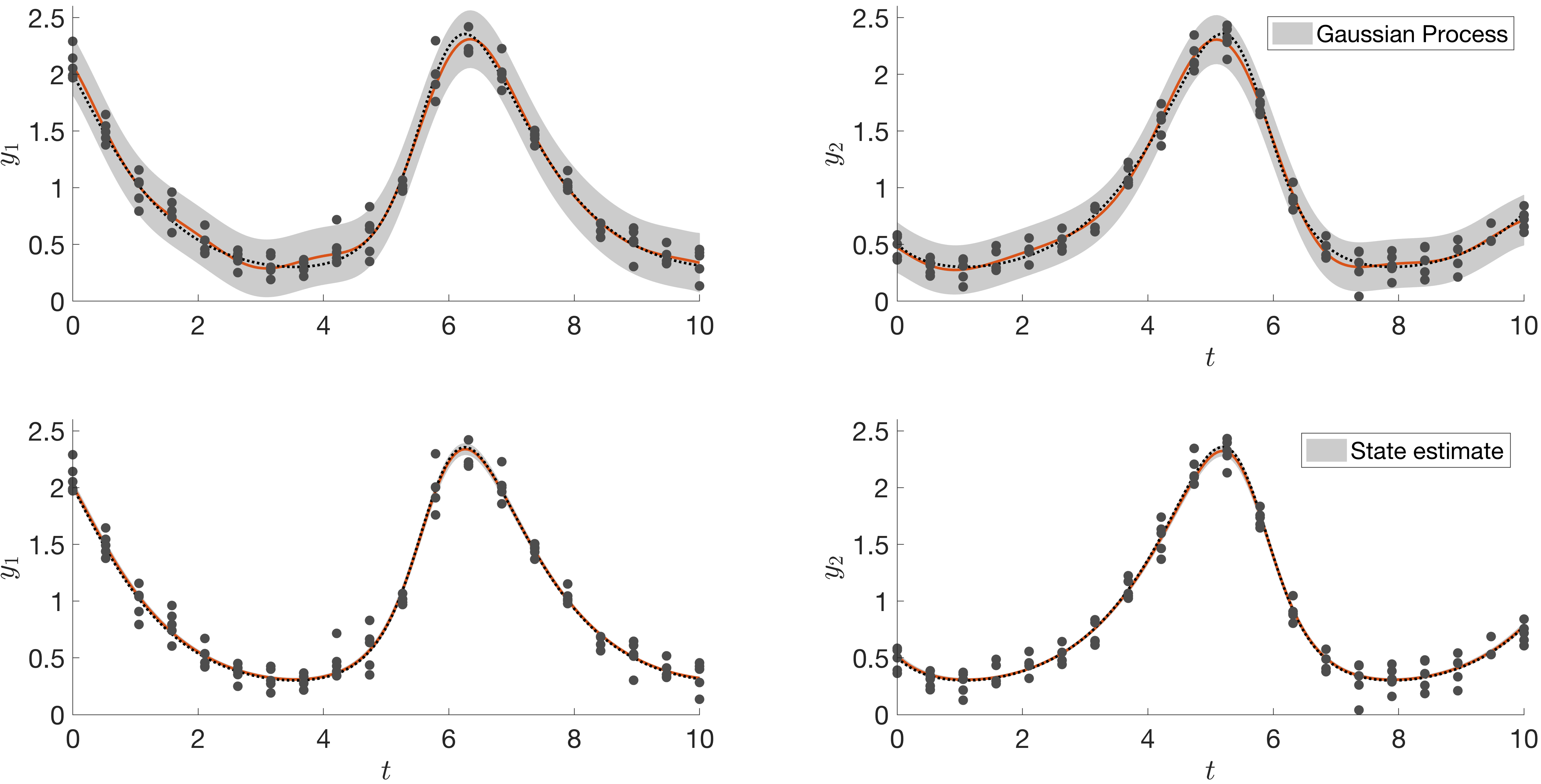}
 \end{center}
 \caption{The top panel shows the dynamics of the true predator and prey system $\bfy_{\rm true}(t)$ in dotted black. Simulated data $\bfd$ are depicted as black dots. The generated GP is represented by its mean $\mu$ (red) and 90\% central confidence region depicted in gray shade. The lower panel shows state estimates after using our framework, again the true predator prey state in dotted black lines. The red line gives the 50th percentile with central 95\% interval shaded in gray.}\label{fig:dataLV}
\end{figure}

We assume that for each state we are given \emph{five} repeated samples at times $t_j$, where the samples are subject to additive noise. Here, $\bfy_{\rm true}(t_j) + \bfvarepsilon$ with $\bfvarepsilon \sim \calN\left(\bfzero,\tfrac{1}{10}\bfI_2\right)$. Hence, $\bfd \in \bbR^{200}$ (see Fig.~\ref{fig:dataLV}). Given these observations, we seek to estimate model parameters $\bfalpha_{\rm true}$ and the initial condition $\bfy_{\rm true}(0)$. Here, $\bfp = [\alpha_1,\alpha_2,y_1(0),y_2(0)]\t$.

The first step is to train the surrogate stochastic process, $\calG$. In this controlled experiment, there is no need to consider heteroskedasticity or censored observations. Thus, we entertain an ordinary GP and use straightforward likelihood-based optimization methods to infer the unknown hyperparameters to a Gaussian covariance kernel. Fig.~\ref{fig:dataLV} provides the mean of the GP predictive surface in red while the gray shaded area reflects the variances. The true curves generating the data are shown as dotted black lines.

Conditional on that fit, we generate 100,000 samples $\left\{\bfg_j(t)\right\}_{j = 1}^{100,000}$ from the predictive equations corresponding to an to the fitted $\calG$ (i.e., we follow the unique-$n$ variation on Eq.~\eqref{eq:gppred}). Numerically, we discretize the sample processes $\bfg_j(t)$ at 201 equidistant times in the interval $[0,10]$ to solve the optimization problem Eq.~\eqref{eq:gpOpt}. For simplicity, we utilize a single shooting method via a direct search method optimizing Eq.~\eqref{eq:gpOpt}, while the dynamical system Eq.~\eqref{eq:lv} is solved via an explicit Runge-Kutta 4 method. Optimization problem Eq.~\eqref{eq:gpOpt} is solved 100,000 times resulting in a set of parameter estimates $\{\widehat \bfp_j \}_{j=1}^{100,000}$. This set $\{\widehat \bfp_j \}_{j=1}^{100,000}$ defines a distribution of estimates of the underlying true parameter values $\bfp_{\rm true}$.
The lower panel of Fig.~\ref{fig:dataLV} illustrates the uncertainty estimates of the states $y_1$ and $y_2$. Note the narrow uncertainty margins and that the true solution lies completely within the error estimates.

\begin{figure}[bthp]
 \begin{center}
  \begin{tabular}{cc}
   \includegraphics[width=0.48\textwidth]{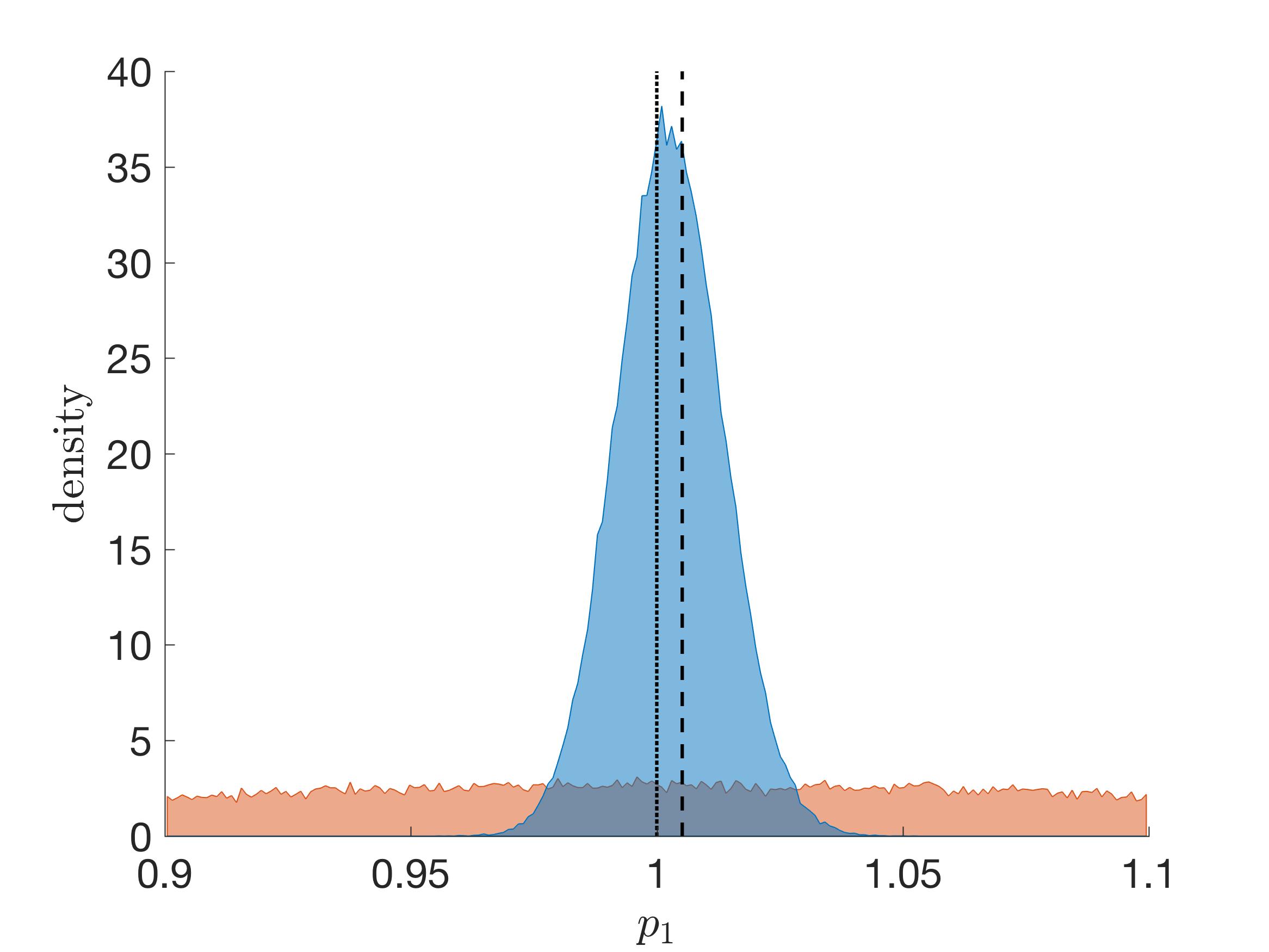} & \includegraphics[width=0.48\textwidth]{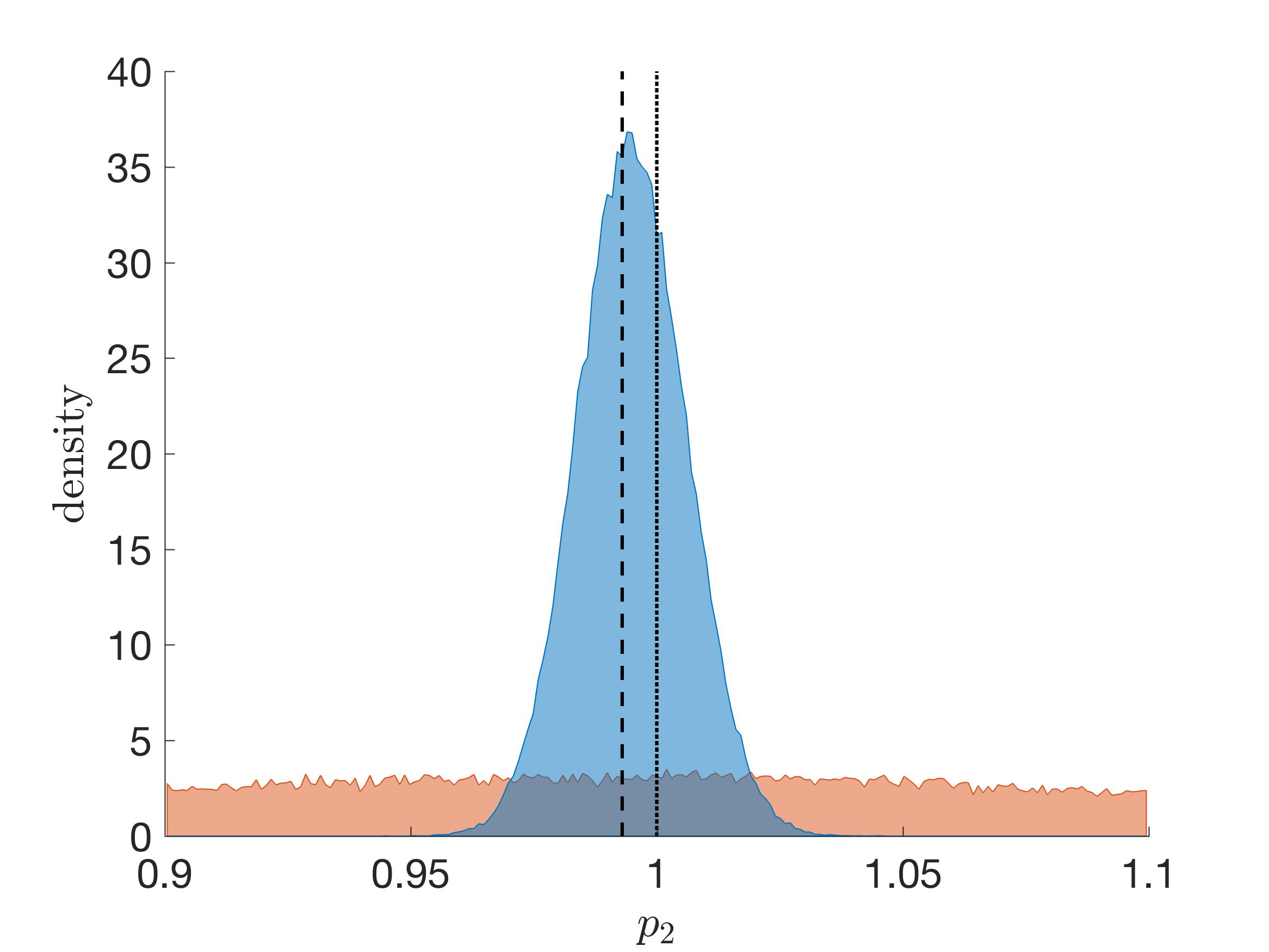} \\
   \includegraphics[width=0.48\textwidth]{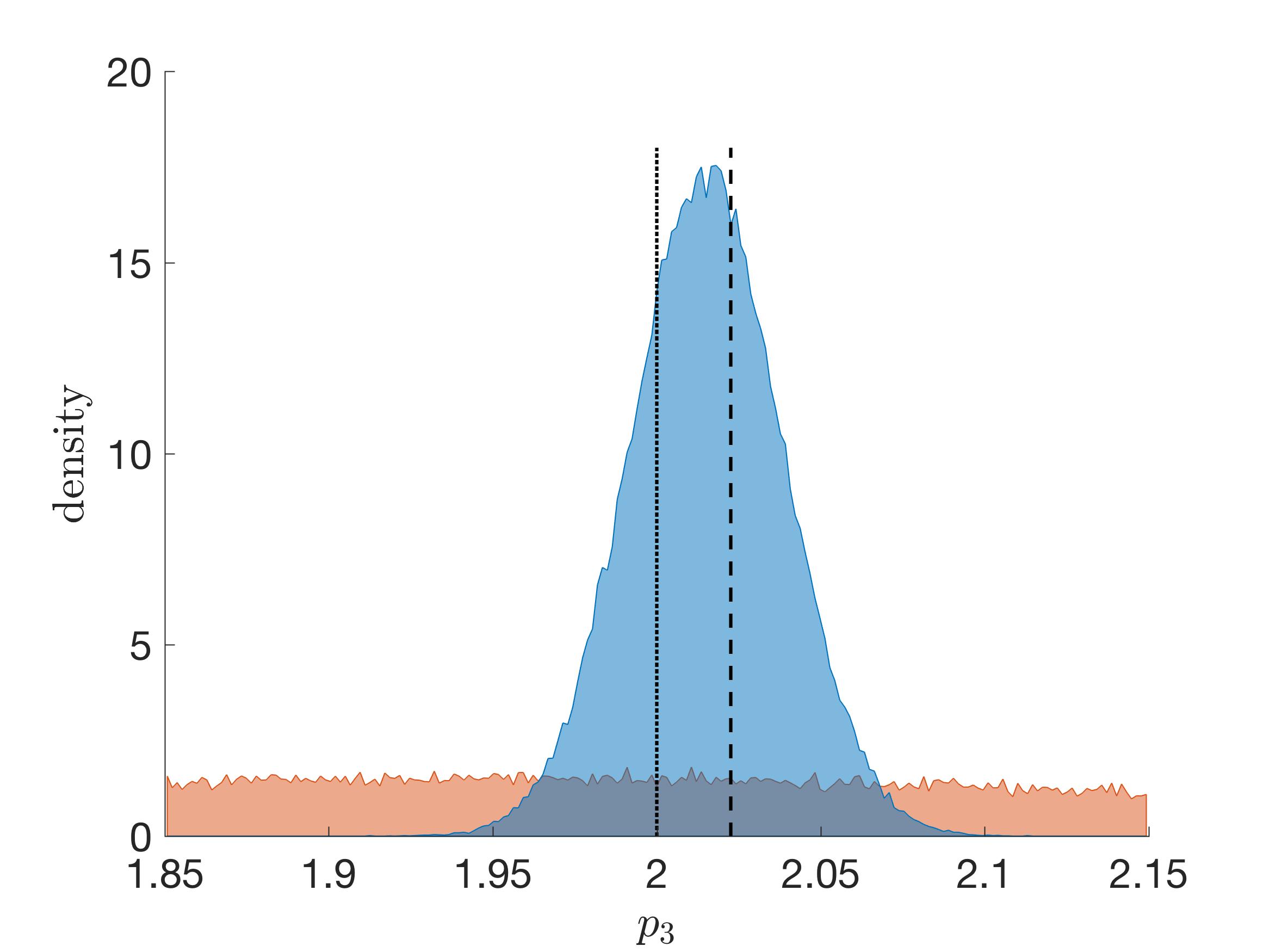} & \includegraphics[width=0.48\textwidth]{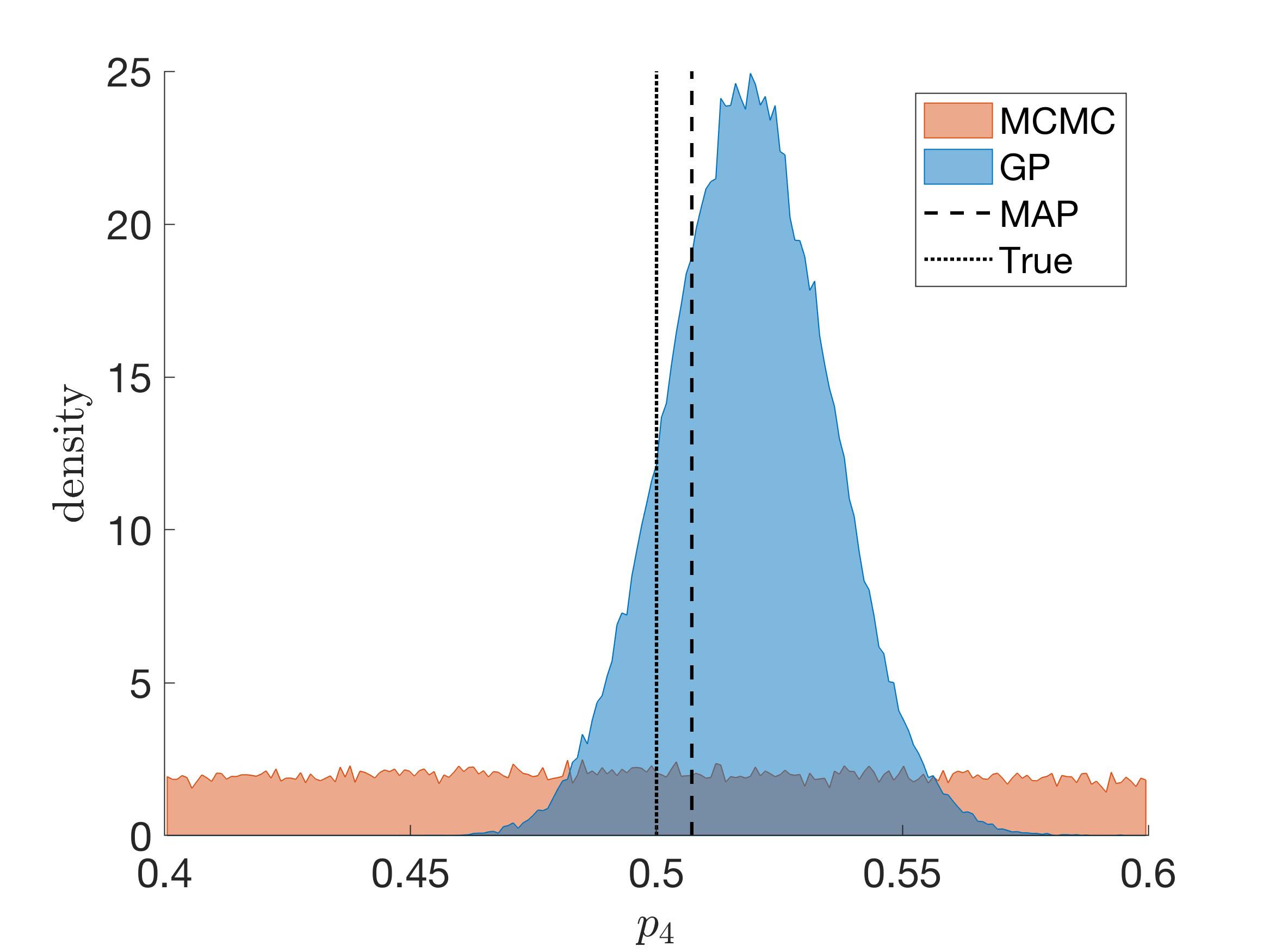}
  \end{tabular}
 \end{center}
 \caption{1-D projected density plots of the estimates $\{\widehat \bfp_j \}_{j=1}^{100,000}$ (GP density in blue) and $\{\tilde \bfp_j \}_{j=1}^{1,000,000}$ (MCMC density in red). The true model parameters $\bfp_{\rm true} = [1, 1, 2, 1/2]\t$ are represented by a dotted line, while the maximum a~posteriori estimate $\bfp_{\rm MAP}$ is represented as a dashed line.} \label{fig:1DLV}
\end{figure}

Fig.~\ref{fig:1DLV} displays the projected 1-D densities for each of the four model parameters $\alpha_1, \alpha_2, y_1(0),$ and $y_2(0)$ (panels 1--4). Highlighted in blue are the densities generated by our new GP based approach while the results for the MCMC alternative \cite{smith2013uncertainty} are overlaid in red. For that method, we chose a uniform prior in the domain $0 \leq p_i \leq 10$ for $i = 1,\ldots, 4$. We initialized the Markov chain at $\bfp_0 = [1, 1, 2, 1/2]\t$ and used random-walk Metropolis proposals as $\calN\left(\tilde\bfp_j, 1/5\,\bfI_4\right)$, where $\tilde\bfp_{j}$ is the previous posterior sample. Mixing in the chain was responsibly good. We determined it to have converged after one million samples and generated another million thereafter to save as posterior draws: $\{\tilde\bfp_j \}_{j=1}^{1,000,000}$. The dotted lines in Fig.~\ref{fig:1DLV} represent the true parameter values $\bfp_{\rm true}$ (dotted line) and the maximum a posteriori (MAP) $\bfp_{\rm MAP} \approx [ 1.005, 0.993, 2.023, 0.507]\t$ (dashed line), respectively. Although our new method generally agrees with MAP obtained via the MCMC, the densities generated by the GPs are narrower and more tightly sandwich the maximum a~posteriori. The tighter densities are due to the regularization implicitly induced by our GP prior, which imposes smoothness and decreases curvature in the state solution $\bfy(t)$.

\subsection{Influenza} \label{sub:numinfluenza}

We next discuss influenza virus parameter estimation and uncertainty quantification as introduced in Section~\ref{sec:bio}, with data visualized in Fig.~\ref{fig:dataInfluenza.jpg} and associated mathematical model detailed by Eq.~\eqref{eq:im}. The data include virus counts, but no data is available for the infected cells $I_1(t)$ and $I_2(t)$ or for the susceptible target cells $T(t)$.

Eq.~\eqref{eq:im} is of the form $ \bfy'= f(t, \bfy, \bfp)$, where the state variable is given by $\bfy(t) = [T(t), I_1(t), I_2(t), V(t)]\t$. We assume that the parameters $\beta, \rho, c, \delta, K_d$ and the initial condition $T(0)$ are unknown, i.e., $\bfp = [\beta, \rho, c, \delta, K_d, T(0)]\t$. We assume that the other parameters and initial condition in Eq.~\eqref{eq:im} are given. Here, we choose $\kappa = 4\ \mathrm{d}^{-1}$ and $I_1(0) = 10\ \mathrm{cells}$, $I_2(0)= 0.02\ \mathrm{cells}$, and $V(0) = 0.07\ \mathrm{TCID}_{50}$. Typically, these are not the initial conditions used in influenza modeling studies (e.g., as. in \cite{smith2018dd}). In particular, there are no productively infected cells ($I_2$) at the time of infection. However, a positive value was necessary for the simulation. The values were chosen arbitrarily.

For the optimization, we use the same setup as in our simulation study of Section~\ref{sub:simstudy}. For 100,000 stochastic processes realizations $\left\{\bfg_j(t)\right\}_{j = 1}^{100,000}$ from data augmented {\tt hetTP}s, we use a single shooting method with direct search optimization and the ODE is solved via a Runge-Kutta 4 method. Numerically, we discretized the residual $\bfs(\bfy(t)) - \bfg_j(t)$ at 3000 equidistant time points. Hence, we generate a set samples $\{\widehat \bfp_j \}_{j=1}^{100,000}$ from the posterior distribution of the underlying true parameter values $\bfp_{\rm true}$.
\begin{figure}[bthp]
 \begin{center}
  \includegraphics[width=\textwidth]{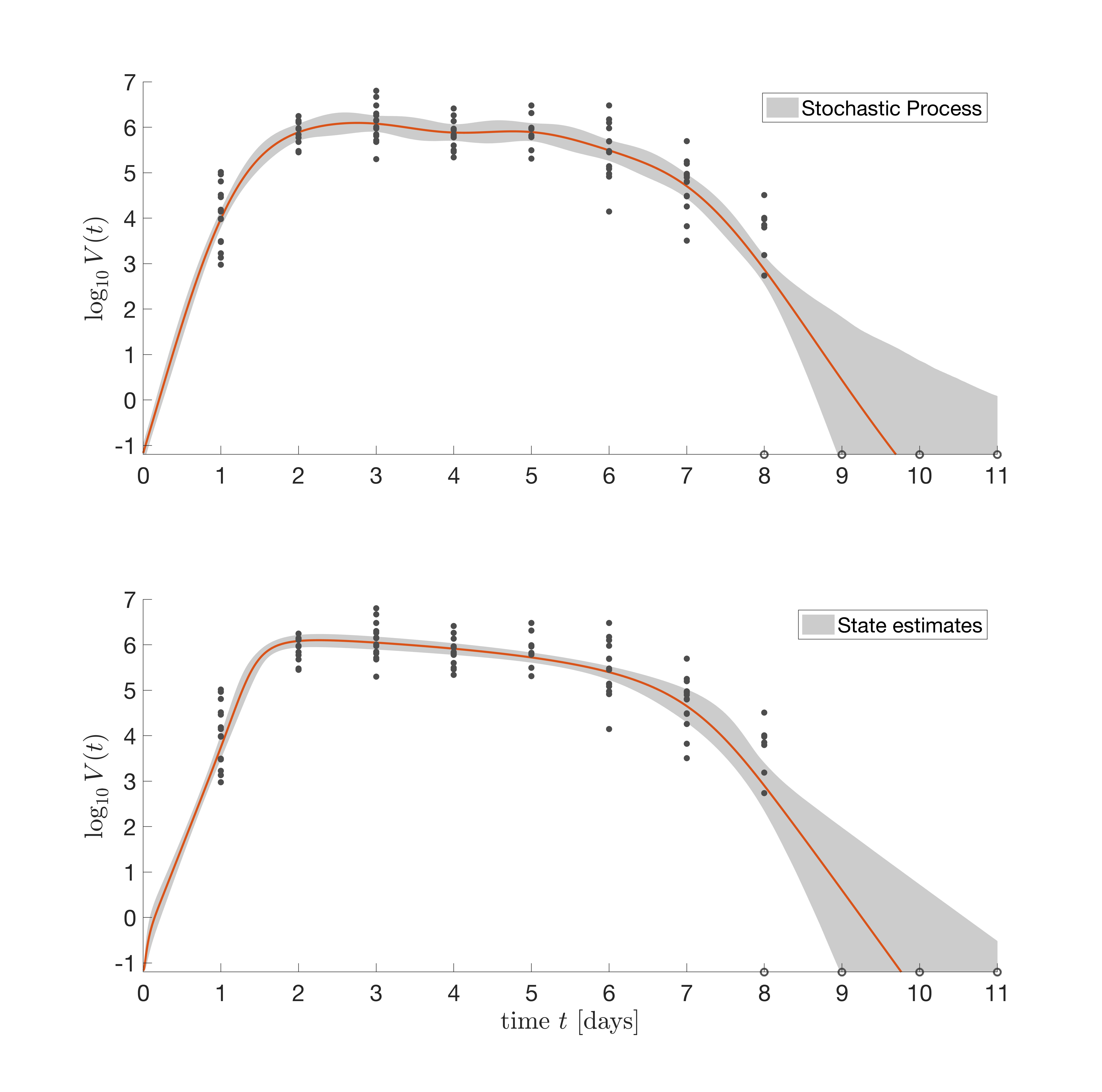}
 \end{center}
 \caption{The top panel shows the statistics of the stochastic process. The lower panel shows the statistics of the reconstructed state solutions of $V$ for the estimated $\left\{ \widehat\bfp\right\}_{j = 1}^{100,000}$ reconstructions. The mean is given represented by red line while the 95 percentiles are shaded in gray, data are given as black dots.}\label{fig:GPandInvSol}
\end{figure}
Fig.~\ref{fig:GPandInvSol} provides an example of the generated data fits. The top panel shows the (de-noised) posterior predictive surface obtained from our fitted {\tt hetTP} surrogate, $\calG$. Sample paths $\bfg_j$ yielding $\widehat{\bfp}_j$ were used to generate a realization of the states derived from the system of differential equations, and the distribution of those curves is shown in the bottom panel of the figure. Notice that the top surface is not strictly unimodal like the bottom surface---as demanded by the ODE. In this way, the figure shows how least-squares calculations ``filter'' posterior inference into parameters of the system of equations via the predictive distribution, as exhibited by their resulting distribution of states.

\begin{figure}[bthp]
 \begin{center}
  \begin{tabular}{cc}
   \includegraphics[width=0.48\textwidth]{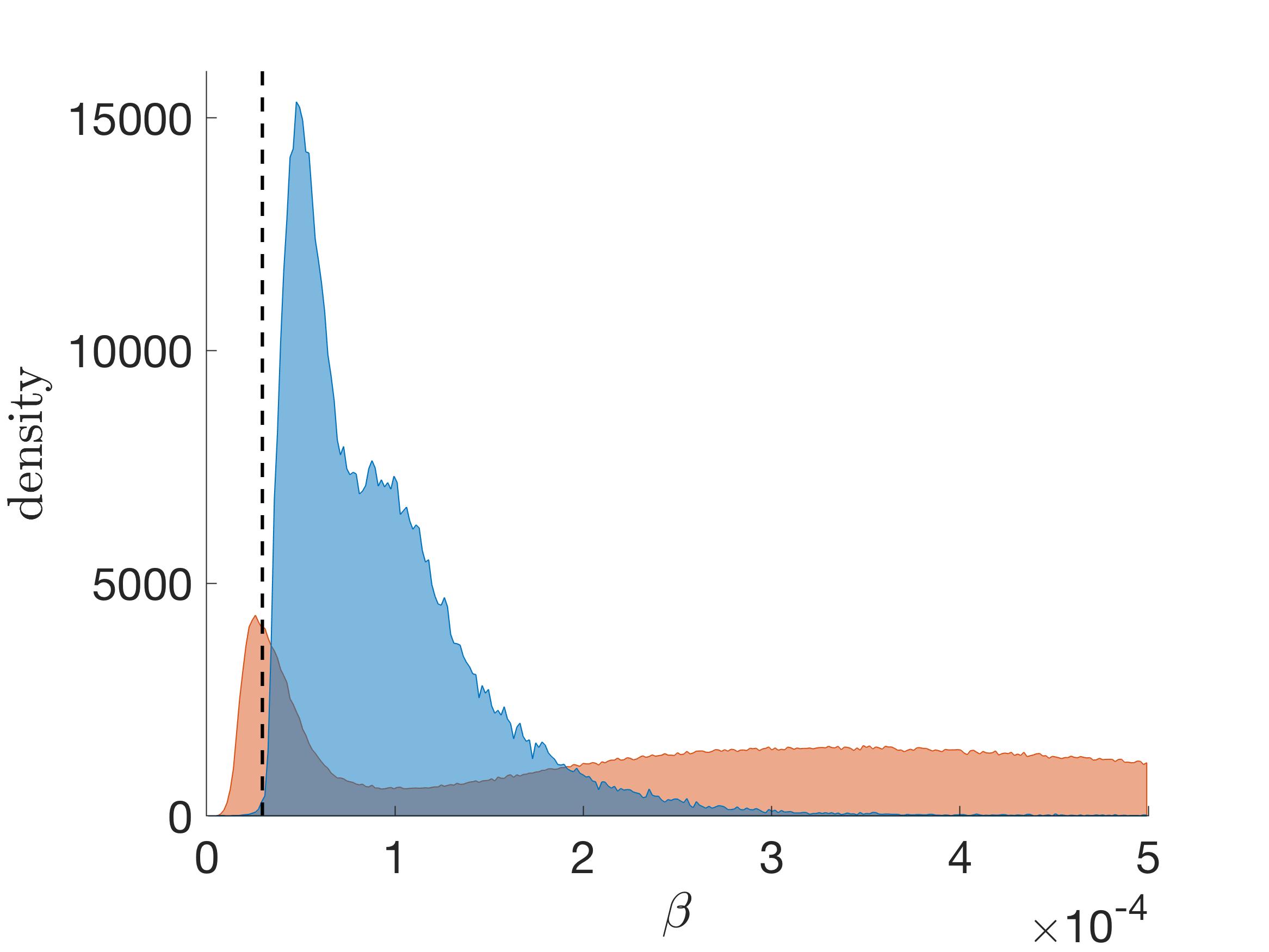} &  \includegraphics[width=0.48\textwidth]{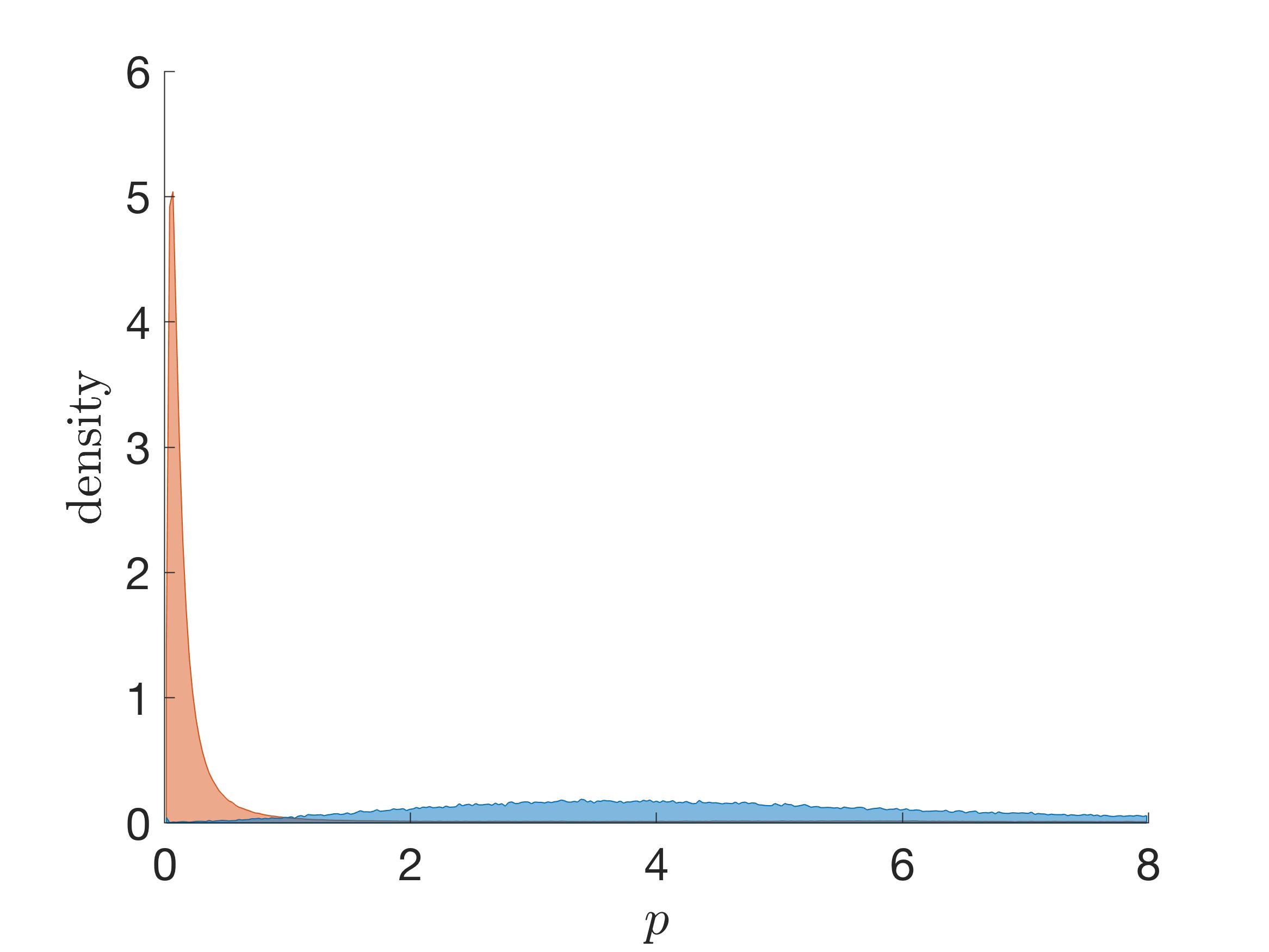} \\
   \includegraphics[width=0.48\textwidth]{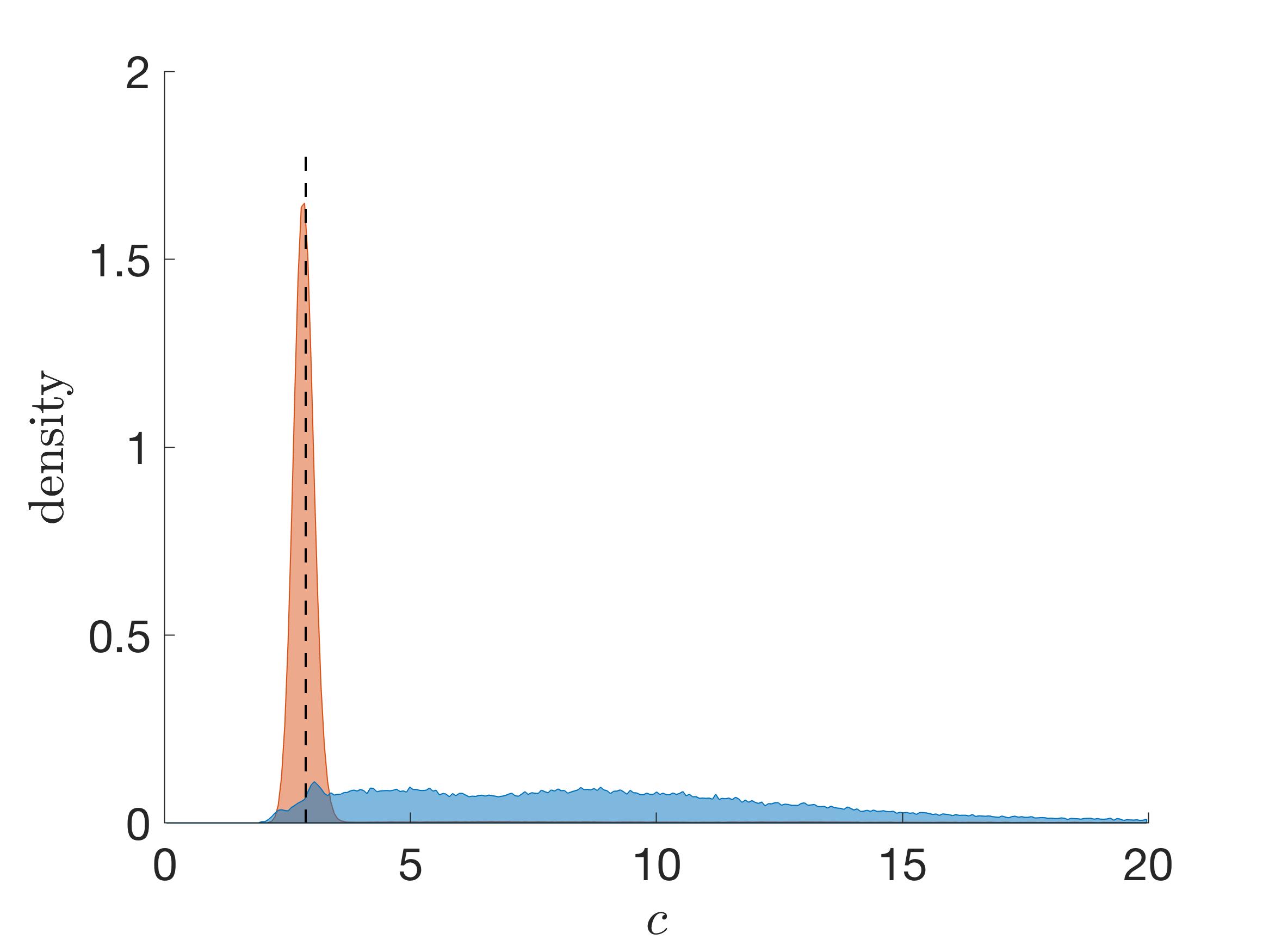} & \includegraphics[width=0.48\textwidth]{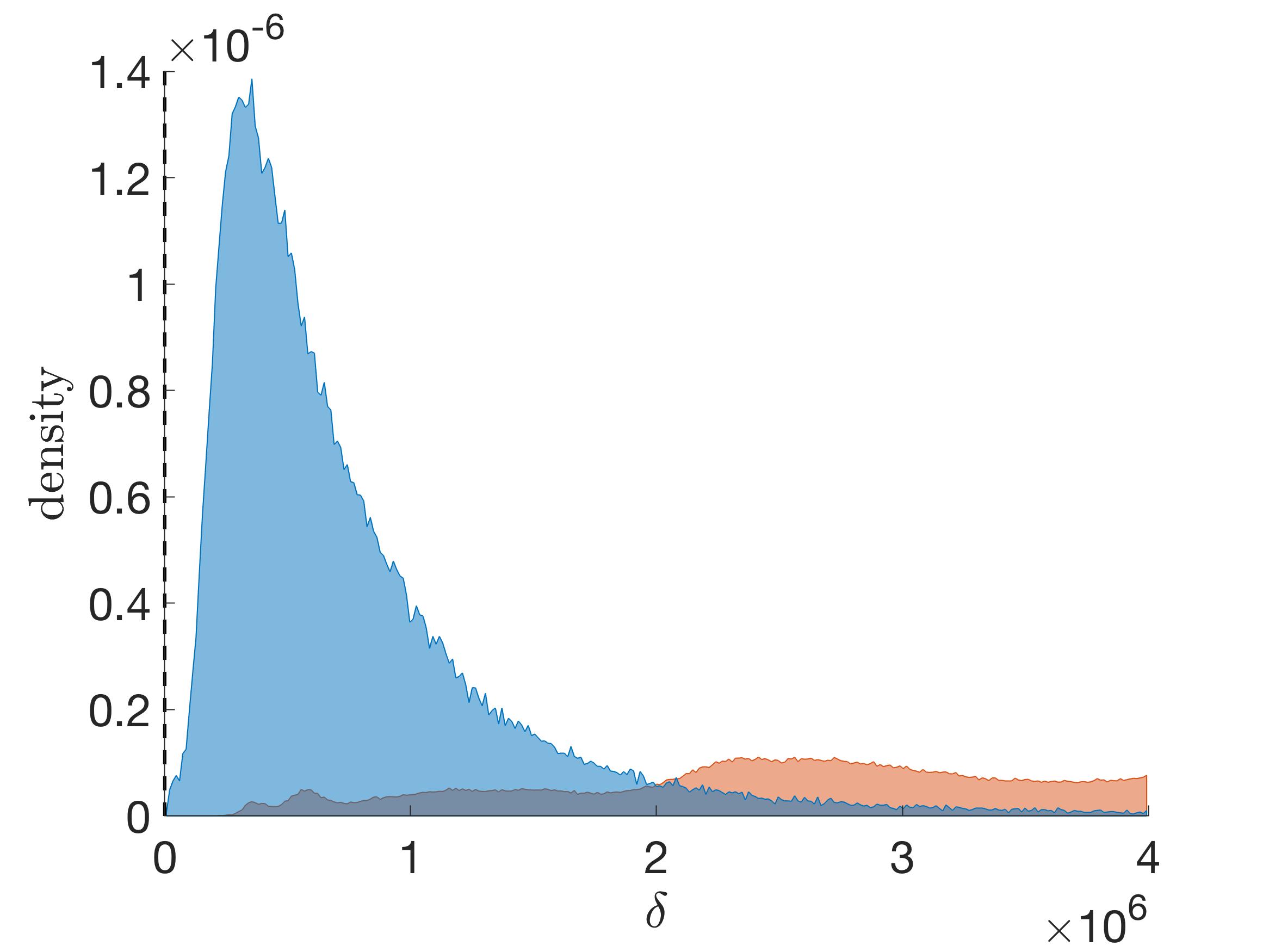} \\
   \includegraphics[width=0.48\textwidth]{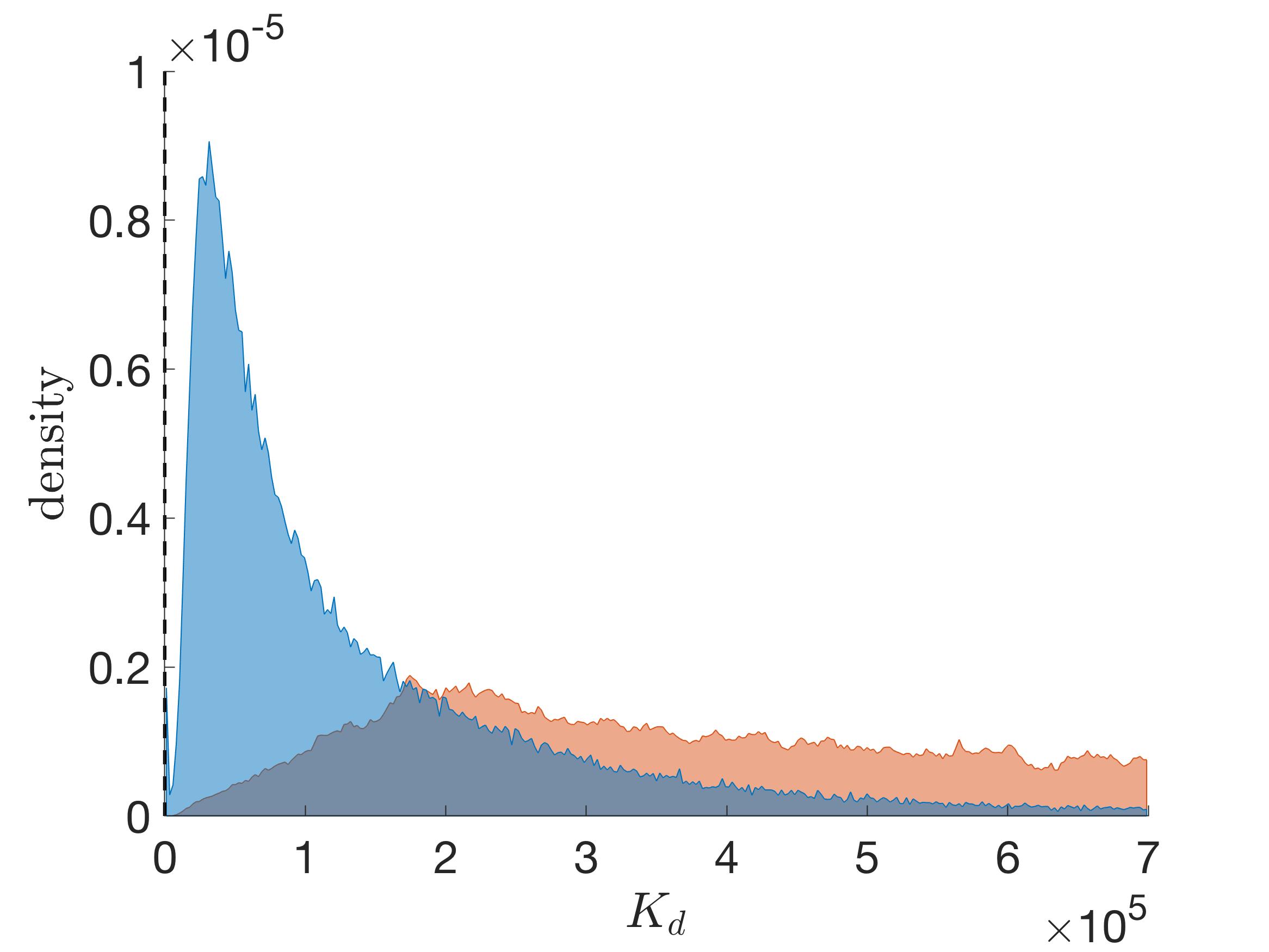} &
   \includegraphics[width=0.48\textwidth]{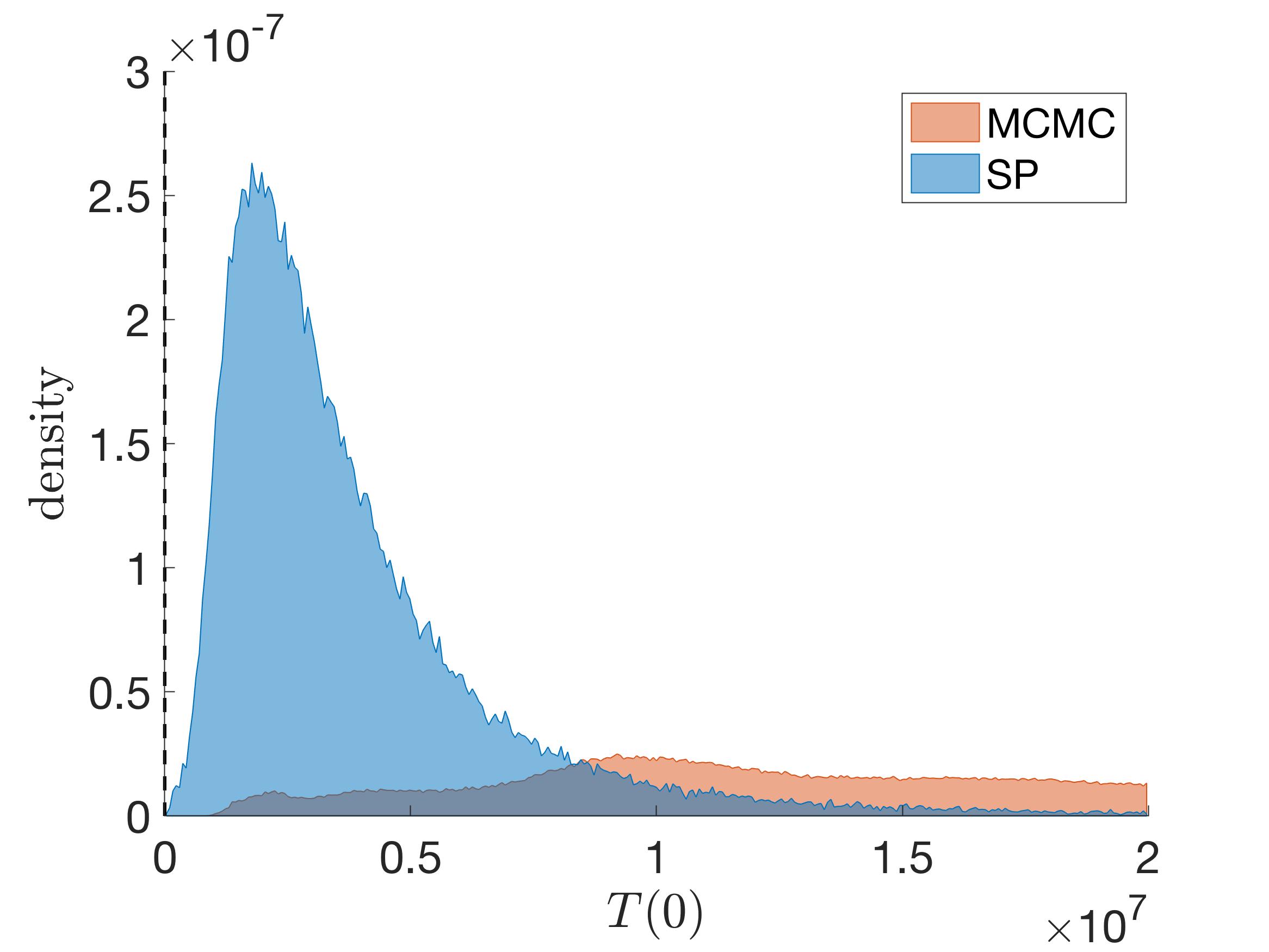}
  \end{tabular}
 \end{center}
 \caption{Projected density plots of the parameter estimates for the influenza model (Eq.~\eqref{eq:im}) and associated data depicted in Fig.~\ref{fig:dataInfluenza.jpg}. Densities generated by the SP are given in blue, while density estimates of the posterior generated by an MCMC chain are in red. The dashed line represent the maximum likelihood estimates. The maximum likelihood estimate of $p$ in panel 2 is omitted because $p_{\rm MLE} \approx 4.4085 \cdot 10^4$.}\label{fig:1DDensityInfluenza}
\end{figure}

Marginal 1-D density results for those posterior distribution on the parameters, the $\widehat{\bfp}_j$, are depicted in Fig.~\ref{fig:1DDensityInfluenza}. Again, the densities utilizing the stochastic process generate a tight distribution while the distribution generated through the MCMC chain give wide uncertainty estimates of the model parameters. A similar MCMC framework for this influenza data was utilized in the master thesis \cite{Torrence2017}. The maximum a posteriori estimate $\bfp_{\rm MAP}\approx[2.9601\cdot 10^{-5}, 4.4085\cdot 10^4, 2.8540, 28.1280, 0.0436, 154.3949]\t$ is given by the black dashed line and exhibits unrealistic estimates due to the ill-posedness in the optimization problem. Again, we draw a comparison to the samples obtained from a simpler posterior via MCMC \cite{smith2013uncertainty}. Uniform priors were chosen in the range(s) $-1 / \eps < p_j < 1 / \eps $, where $\eps$ is given by the machine precision, and random-walk Metropolis proposals were generated as $\calN\left(\tilde \bfp_{j}, 1/5\, \bfI_6\right)$. The maximum likelihood estimate was used to generate a starting value. We determined the MCMC chain to have converged after one million samples and the next one million were saved as posterior samples $\{\tilde \bfp_j \}_{j=1}^{1,000,000}$.

The parameter distributions shown in Fig.~\ref{fig:1DDensityInfluenza} illustrate the improvement made by the SP method compared with the MCMC method. In addition, the results from fitting via SP more closely reflect those obtained from using traditional global optimization methods (e.g., adaptive simulated annealing) \cite{smith2018dd}, which have been shown to yield accurate estimates \cite{smith2013kinetics,smith2016critical}. Altering the data through censoring does skew the values of the parameters, which is expected. The effect is particularly evident in the value of $\delta$, which dictates the viral decay dynamics and is the most sensitive parameter \cite{smith2010aat,smith2011effect,smith2018dd}. Here, we assumed that the censored data decreased monotonically, which is unconventional in viral kinetic modeling. Most often, censored data is imputed as one half the LOD and without any dynamical restrictions. By imposing monotonicity across several times with repeated LOD measurements rather than truncating at the first instance (i.e., 8--11 d pi versus 8--9 d pi), solutions inconsistent with experimental observations were produced (Fig.~\ref{fig:GPandInvSol}). That is, the resulting model trajectories do not capture the rapid viral clearance in some mice between 7--8 d pi, where the values may indeed be true zeros. In addition, they suggest that animals may have viral loads above the LOD at 9 d pi and not clear the infection until 11 d pi, neither of which have been observed. Thus, one should use caution when censoring data as biological inference can be inhibited. However, the consistency in parameter distributions and behavior (e.g., correlation between $\rho$ and $c$) between the results here and the results in Smith et al.~\cite{smith2018dd} support the accuracy of the fitting method.

\section{Conclusion \& Discussion} \label{sec:discussion} Our proposed methods consists of two parts. In the first phase, we fit a surrogate stochastic process to given data. In the second phase, we use a set of samples from the posterior predictive distribution of that fitted stochastic process to generate surrogate data, which are then ``passed through'' a typical scheme used to tune a mathematical model's parameterization to the data. Such an approach represents a novel means of obtaining a posterior distribution on model parameters. A major advantage of this procedure is that we can induce prior knowledge (e.g., smoothness of the states) directly into the process samples, and handle other nuances in the data like input-dependent noise, leptokurtic errors, and censoring. The result is a far more ``focused'' posterior distribution compared to other Bayesian alternatives. A further advantage is that our method provides uncertainty estimates, but does not rely on the Markov property as in MCMC methods. Hence, this method is embarrassingly parallelizable. Depending on the nature of other, similar applications, we envision many opportunities for extension via adaptations to the prior implied by the chosen family of surrogate stochastic processes, e.g., to deal with large amounts of data or additional known features in the data (e.g., symmetries).

Although our problem setting has much in common with those typically tackled within the Kennedy \& O'Hagan \cite{kennedy:ohagan:2001} (KOH) framework or related setups \cite{higdon2004combining}, there are several important reasons why those approaches are not well suited for our setting. One has to due with Bayesian computation. Inference in KOH settings require MCMC with likelihood evaluations. This incurs a cubic cost (in the number of data points) for evaluation. That through exploration of the posterior computationally cumbersome in moderate data size settings, and/or in parameters spaces of moderate size. Parallelization offers no respite due to the inherently serial nature of the Metropolis steps typically involved. Another has to due with incorporating known dynamics into the prior on the surrogate stochastic process, and related issues in handling censored observations. It is fairly easy to implement a rejection sampling scheme, such as the one described in \ref{sec:censor}, to generate appropriately constrained realizations from the posterior predictive distribution and subsequently map those (deterministically) to parameter values. It is quite another to approach the problem from the other direction by accepting or rejecting parameter settings that make such surfaces more or less likely under the posterior distribution. Because it was not obvious how we could accommodate these constraints in a KOH framework, we found it difficult to entertain it as a comparator in our empirical work.

The main focus of this work was to provide a proof of concept of our new methods. Many extensions, modification, and analysis remain open and will be subject to future research. For instance, the main computational burden of our proposed method is the repeated (but parallel) optimization procedure. Solving these ODE constrained optimization problems may be done more efficiently by using Newton type optimization methods coupled with efficient ODE solvers and informed initial parameter guesses as proposed in \cite{chung2015robust}. Our methods can also be extended to optimal experimental design problems with underlying ODE systems \cite{Chung2012}. Further investigations will also be directed towards loosening our stringent statistical assumption on the data: we assumed that the data comes from a single underlying true parameter. This is an oversimplification and future research may target distributions of true parameter.

\bibliographystyle{siamplain}
\bibliography{combinedRef.bib}

\appendix
\section{Derivatives of log-likelihood for Student-t processes}
\label{ap:TP}

Recall that the full-$N$ log-likelihood is given by
\begin{equation*}
\log(L) = -\frac{N}{2} \log((\alpha - 2) \pi) - \frac{1}{2}\log|\KN| + \log \left(\frac{\Gamma \left(\frac{\alpha+N}{2}\right)}{\Gamma \left(\frac{\alpha}{2} \right)} \right) - \frac{(\alpha + N)}{2} \log \left( 1 + \frac{\beta}{\alpha - 2} \right) .
\end{equation*}

By taking into account savings from replicates, the reduced unique-$n$ log-likelihood is:

\begin{align*}
\log(L) =& -\frac{N}{2} \log((\alpha - 2) \pi) -  \frac{1}{2} \log |\tau^2 \Cn + \An^{-1}\Sn|  - \frac{1}{2} \sum\limits_{i=1}^n \left[(a_i - 1)\log \lambda_i + \log a_i \right]\\
 &+ \log \left(\frac{\Gamma \left(\frac{\alpha+N}{2}\right)}{\Gamma \left(\frac{\alpha}{2} \right)} \right) - \frac{(\alpha + N)}{2} \log \left( 1 + \frac{\beta}{\alpha - 2} \right),
\end{align*}
with $\beta = \bfd^\top \SN^{-1} \bfd - \bar \bfd^\top \An \Sn^{-1}\bar \bfd + \bar \bfd^\top (\tau^2 \Cn + \An^{-1} \Sn)^{-1} \bar \bfd$.\\

For likelihood based optimization of the hyperparameters, derivatives become very useful. Shah et al. in
\cite{shah:wilson:ghahramani:2014} provide derivatives with respect to $\alpha$ and $\theta$, that we complement for our setup. The derivative with respect to $\alpha$ is

\begin{align*}
\frac{\partial }{\partial \alpha} \log L = -\frac{N}{2(\alpha - 2)} + \frac{1}{2} \psi\left(\frac{\alpha + N}{2} \right) - \frac{1}{2} \psi\left(\frac{\alpha}{2} \right)  - \frac{1}{2} \left( 1 + \frac{\beta}{\alpha - 2} \right) \\
+ \frac{(\alpha + N) \beta}{2(\alpha -2)^2 + 2\beta (\alpha - 2)},
\end{align*}
with $\psi$ the digamma function.\\

For the other hyperparameters, denoting $\Un = \tau^2 \Cn + \An^{-1}\Sn$:
\begin{equation*}
\frac{\partial }{\partial \cdot} \log L = -\frac{1}{2} tr \left(\Un^{-1} \frac{\partial \Un}{\partial \cdot} \right)
 -\frac{\alpha + N}{2(\alpha + \beta - 2)} \frac{\partial \beta}{\partial \cdot} -\frac{1}{2} \sum \limits_{i = 1}^n (a_i - 1) \frac{\partial \log \lambda_i}{\partial \cdot},
\end{equation*}

in particular we get
\begin{equation*}
\frac{\partial }{\partial \theta} \log L = -\frac{\tau^2}{2} tr \left(\Un^{-1} \frac{\partial \Cn}{\partial \theta} \right)
 + \frac{\alpha + N}{2(\alpha + \beta - 2)} \tau^2 \bar \bfd^\top \Un^{-1} \frac{\partial \Cn}{\partial \theta} \Un^{-1} \bar \bfd
\end{equation*}

\begin{equation*}
\frac{\partial }{\partial \tau^2} \log L = -\frac{1}{2} tr \left(\Un^{-1} \Cn \right)
 + \frac{\alpha + N}{2(\alpha + \beta - 2)} \bar \bfd^\top \Un^{-1}  \Cn \Un^{-1} \bar \bfd
\end{equation*}

\begin{align*}
\frac{\partial}{\partial \Lan} \log L = - \frac{1}{2}\An^{-1} \Diag(\Un^{-1}) - \frac{\An - \I_n}{2} \Lan^{-1}\\
 + \frac{\alpha + N}{2(\alpha + \beta - 2)} \An \mathbf{S} \Lan^{-2} + \An^{-1} \Diag(\Un^{-1} \bar \bfd)^2
\end{align*}


\end{document}